\pdfoutput=1
\documentclass[PAPER, atlasdraft=false, UKenglish, cernpreprint, texmf, orcidlogo]{atlasdoc}
\usepackage[backend=biber, block=space, subcaption, rotating, longtable, multirow, subcaption]{atlaspackage}
\usepackage[italic]{hepnicenames}
\usepackage{adjustbox}
\usepackage{enumerate}
\usepackage{array}
\usepackage{listings}
\usepackage{multirow}
\usepackage{tabularx}
\usepackage{amsmath}
\usepackage{xcolor}
\usepackage{hhline}
\usepackage{cancel}
\usepackage{tikz}
\usepackage{xfrac}
\usetikzlibrary{shapes.geometric, arrows}
\usepackage{siunitx}
\newcolumntype{~}{>{\global\let\currentrowstyle\relax}}
\newcolumntype{^}{>{\currentrowstyle}}
\newcommand{\rowstyle}[1]{\gdef\currentrowstyle{#1}%
#1\ignorespaces
}

\usepackage{atlasbiblatex}

\usepackage[hepparticle, hepprocess, snippets, xref=false, unit=true]{atlasphysics}
\graphicspath{{logos/}{figures/}}

\usepackage{ANA-TOPQ-2020-17-PAPER-defs}
\usepackage[normalem]{ulem}

\AtlasTitle{Constraints on  effective field theories
via quadruple-differential angular decay
rates from $t$-channel single-top-quark production
at $\sqrt{s}=13$~TeV with the ATLAS detector}

\AtlasVersion{3.1}

\AtlasAbstract{
Events with $t$-channel single top quarks are used to probe effective field theory operators
in $\sqrt{s}=13$~TeV proton--proton collision data corresponding to $140$~fb$^{-1}$ recorded
by the ATLAS detector at the Large Hadron Collider. To extract a maximum amount of
information from these events, the joint fourfold probability distribution for four
kinematic angles is simultaneously analysed. A method leveraging Fourier techniques to handle
the challenges arising from this approach is employed. The relevant effective
field theory operators are those sensitive to top-quark decay and $t$-channel
production vertices. Their Wilson coefficients are tightly constrained, with
results compatible with Standard Model predictions.
}

\AtlasRefCode{TOPQ-2020-17}

\PreprintIdNumber{CERN-EP-2025-208}

\AtlasJournal{JHEP}
\AtlasJournalRef{\JHEP 07 (2026) 107}
\AtlasDOI{10.1007/JHEP07(2026)107}

\AtlasCoverEgroupAnalysisTeam{atlas-topq-2020-17-atlas-analysis-team@cern.ch}

\hypersetup{pdftitle={ATLAS document},pdfauthor={The ATLAS Collaboration}}

\begin{document}

\maketitle

\tableofcontents

\section{Introduction}
%

% The next lines are included from the .//sections/introduction.tex input file
%
%
%
%
%
%

An interesting avenue in searching for possible interactions beyond the Standard Model (BSM)
is looking for non-zero Wilson coefficients (WCs) of effective field
theories (EFTs), and in particular
those that couple to the top quark.
Owing to its heavy mass, $\mt \approx \SI{172.5}{\GeV}$~\cite{TOPQ-2019-13}, and short lifetime,
$1/\Gamma_t \sim{\cal O}(10^{-25})$~s~\cite{ParticleDataGroup:2024cfk}, the top quark decays
before hadronising, mainly through the weak process $t \rightarrow Wb$~\cite{ParticleDataGroup:2024cfk}.
This makes the measurement of its production and decay kinematics a unique probe of BSM physics.

In the bottom-up approach, an effective theory is an approximation to
physics at a higher energy scale. A prominent example is the Standard
Model effective field theory (SMEFT)~\cite{Brivio:2017vri}. In the
SMEFT, the Standard Model (SM) is treated as the EFT of an unknown BSM
theory at a higher scale, where the scale of the SM is characterised
by the electroweak (EW) scale given by the vacuum expectation value
$v$. Even though the BSM theory is unknown, its effects can still be
parameterised by higher-dimensional operators in a power expansion. Therefore, the effective SM Lagrangian can be written as an expansion
\begin{equation*}
\mathcal{L}_{\textnormal{eff}} = \mathcal{L}_{\textnormal{SM}}^{(4)} + \frac{1}{\Lambda} \sum_{i} c_i^{(5)} \mathcal{O}_i^{(5)} + \frac{1}{\Lambda^2} \sum_{i} c_i^{(6)} \mathcal{O}_i^{(6)} + \mathcal{O}\left(\frac{1}{\Lambda^3} \right)+...,
\end{equation*}%
where the $c_i$ are the WCs~\cite{Wilson:1972ee} of the operators
$\mathcal{O}_i$, containing all degrees of freedom below the
scale $\Lambda$, which is the energy scale of the BSM physics.
The scale $\Lambda$ is the cut-off scale at which the
power expansion breaks down and the BSM theory cannot be explained by
the effective theory anymore.
At this energy scale, the heavy particles that were parameterised
out in the EFT approach can instead be produced directly (`on-shell'),
and their dynamics must be explicitly included in the theory.
The number in parentheses is the energy dimension of the
respective operators. Usually, $\mathcal{O}^{(5)}$ terms are not considered because they violate baryon-number conservation.

BSM physics at higher scales could manifest itself at lower energies through its effective interactions, leading to detectable deviations from SM predictions~\cite{Willenbrock:2014bja}. There are a large number of top-quark-related dimension-six operators that can be constructed from the SM degrees of freedom~\cite{Buchmuller:1985jz}.
A complete, non-redundant basis expressed in the Warsaw basis was given in Ref.~\cite{Grzadkowski:2010es} and consists of 59 independent operators, not counting flavour indices and Hermitian conjugates.
Among the 59 operators,
this analysis focuses solely on the five operators contributing to the
\tch single-top-quark production process or the $\Pqt\ \to \PW\Pqb$ decay, out of which three have complex WCs.

\begin{figure}[!htb]
\centering
{\includegraphics[width=2.5in]{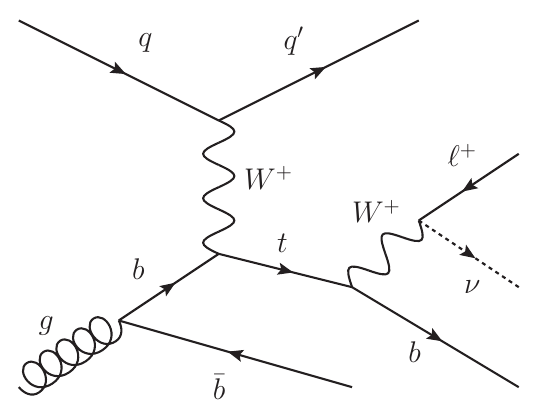}}
\caption{Representative LO Feynman diagram for \tch\ single-top-quark production and decay in the 4FS. Here $q$ represents a $\Pqu/\Pqc$ or
$\Paqd/\Paqs$ quark, and $q^\prime$ represents
a $\Pqd/\Pqs$ or $\Paqu/\Paqc$ quark. The initial
\bquark typically arises from a gluon splitting into a \bbbar pair.}
\label{fig:tchannel}
\end{figure}

In proton--proton (\pp) collisions at the Large Hadron Collider (LHC)~\cite{Evans:2008zzb}, the \tch\ has the largest
rate among all single-top-quark production channels~\cite{CMS-TOP-17-011,TOPQ-2019-02}.
Moreover, it is the only channel that produces a polarised top quark, providing unique access
to the imaginary sector of the Lagrangian through analyses of differential angular decay rates.
\Cref{fig:tchannel} shows
a representative Feynman diagram for \tch\ single-top-quark production, occurring at leading order (LO) in the four-flavour factorisation scheme (4FS), but at next-to-leading order (NLO) in the five-flavour scheme (5FS)~\cite{PhysRevLett.102.182003,Maltoni:2012pa,Bothmann:2017jfv}.
In this diagram, a light-flavour quark $q$ (i.e. \Pqu-, \Pqc-, \Paqd-, or \Paqs-quark) from one of the colliding protons
interacts with a gluon (splitting into a \bbbar pair) by exchanging a space-like virtual \PW boson,
producing a top quark \Pqt and a recoiling light-flavour quark $q'$, the
so-called spectator quark.
The second \bquark from gluon scattering is expected to have a softer
transverse momentum (\pT) spectrum and a broader pseudorapidity ($\eta$) distribution
than the \bquark produced in the top-quark decay.
It is generally either not detected in the experiment or does not pass
the event selection.
The final state includes one light-flavour spectator jet, one \bjet, one lepton and missing transverse momentum from the neutrino.%

\begin{figure}[!htb]
\centering
{\includegraphics[width=4in]{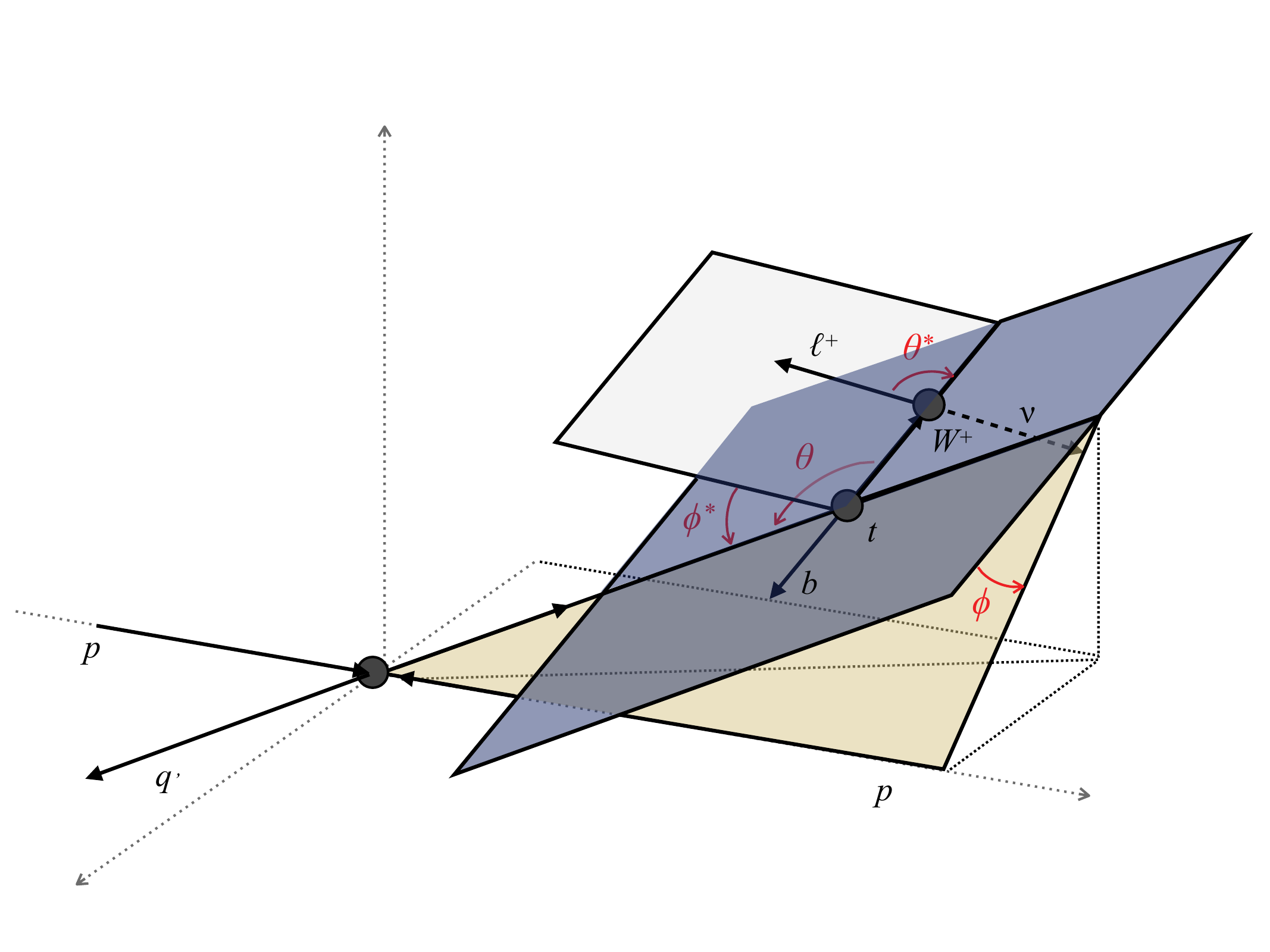}}
\caption{Illustration of the definition of the four angles
$(\theta, \phi, \theta^*, \phi^*)$ in the $t\rightarrow
Wb\rightarrow \Pl \nu b$ decay chain. The incoming protons,
the spectator, top and \Pqb quarks, the \PW boson, and the
lepton and neutrino are denoted by $p$, $q'$, \Pqt, \Pqb, \PW,
\Pl, and $\nu$, respectively. The angles $\theta$ and $\phi$
specify the direction of the \PW-boson (or spectator-quark)
momentum in the top-quark rest frame, while $\theta^*$ and $\phi^*$ specify the direction
of the charged lepton momentum in the \PW-boson rest frame.}
\label{fig:four_angle_definitions}
\end{figure}

The differential distribution of the top-quark decay, $\Pqt \rightarrow \PW \Pqb \rightarrow \Pl \nu \Pqb$,
can be fully determined by four angles~\cite{Aguilar-Saavedra:2017wpl}. The first two
angles $(\theta, \phi)$ are the spherical coordinates of the outgoing
\PW-boson momentum direction in the top-quark rest frame, defined in a Cartesian coordinate system $(x, y, z)$.
The frame is chosen by first selecting the proton beam whose \pz matches that of the outgoing spectator jet, aligned with the incoming proton beam.
Then, following Ref.~\cite{Aguilar-Saavedra:2014eqa}, three orthogonal
directions are defined.
The $\hat{z}$ direction is the direction of the momentum of the
spectator jet, $\vec{p}_s$, in the top-quark rest frame. The $\hat{y}$
direction is taken to lie along $\hat{z}\times\hat{p}_q$, where
$\vec{p}_q$ is the direction of the incoming proton, in the
top-quark reference frame.
The $\hat{x}$-direction is then chosen so as to make a right-handed coordinate system.
The remaining
two angles $(\theta^*, \phi^*)$ are the spherical coordinates of the charged lepton's
momentum in the \PW-boson rest frame, in the coordinate system $(x', y', z')$ obtained
by a \enquote{standard} boost from the previous frame, such that the $\hat{z}'$ axis is the
direction $(\theta, \phi)$ of the \PW-boson momentum and the $\hat{y}'=\hat{z}'\times\hat{p}_s$ axis lies in the
$x{-}y$ plane. %
In this system, $\phi$ is the angle between the production plane and the top-quark decay frame, while $\phi^*$ is the angle between the top-quark decay plane and the \PW-boson decay frame.
The four angles are illustrated in~\cref{fig:four_angle_definitions}.

\FloatBarrier

Many previous measurements by ATLAS~\cite{TOPQ-2018-10,TOPQ-2020-20,TOPQ-2018-24,TOPQ-2019-02}
and CMS~\cite{CMS-TOP-17-020,CMS-TOP-19-001,CMS-TOP-21-001,CMS-TOP-22-006,CMS-TOP-21-003} provide
constraints on the WCs examined in this paper. %
In addition, other measurements~\cite{TOPQ-2013-12,TOPQ-2014-17, TOPQ-2015-06,TOPQ-2015-11} by ATLAS constrain
BSM contributions to the \Wtb\ vertex, albeit
in the language of anomalous couplings rather than EFT coefficients.
Constraints on the real parts of the three complex WCs examined in this paper have also been obtained from an ATLAS measurement of
\PW-boson helicity in top-quark decay~\cite{TOPQ-2018-02}, although there were
not enough independent observables available to constrain all of the WCs simultaneously.

This paper is structured as follows.
The WCs chosen for study in this analysis are introduced in \cref{sec:motivation}.
A brief description of the ATLAS detector is given in \cref{sec:atlas_detector}. Details about the data samples, the Monte
Carlo (MC) simulation event samples, and the background estimation are provided in \cref{sec:data_mc_samples}.
Then the object definitions, reconstruction, and the event selection are discussed
in \cref{sec:object} and \cref{sec:event_selection}, respectively. \cref{sec:systematics} describes the
systematic uncertainties arising in this analysis.
\cref{sec:analysis} presents the
analysis strategy, including the Fourier techniques employed to extract observables from the dataset, the manner in which they are related to the WCs, and the construction of the profile-likelihood. Finally, the results are discussed
in \cref{sec:result} and the conclusions are given in \cref{sec:conclusion}.

% End of text imported from the .//sections/introduction.tex input file

\section{Choice of Wilson coefficients}
\label{sec:motivation}

% The next lines are included from the .//sections/motivation.tex input file
%
%

%
%
%
%
%
%
%
%
%
%
%
Written in the physical basis, where the Higgs field of the Warsaw basis is set to its vacuum expectation value~\cite{Grzadkowski_2008}, four of the operators contribute to the \Wtb\ Lagrangian as
\begin{align*}
\mathcal{O}_{\varphi Q}^{3} & \supset -\frac{g v^2}{\sqrt{2}} \bar{b}_\text{L} \gamma^{\mu} t_\text{L} W_{\mu}^-, \nonumber\\
\mathcal{O}_{\varphi tb} & \supset \left[ \frac{g v^2}{2\sqrt{2}} \bar{b}_\text{R} \gamma^{\mu} t_\text{R} W_{\mu}^- \right]^{\dagger}, \nonumber\\
\mathcal{O}_{tW} & \supset 2 v \bar{b}_\text{L} i \sigma^{\mu\nu} q_{\nu} t_\text{R} W_{\mu}^-, \nonumber\\
\mathcal{O}_{bW} & \supset \left[ 2 v \bar{b}_\text{R} i \sigma^{\mu\nu} q_{\nu} t_\text{L} W_{\mu}^- \right]^{\dagger},
\end{align*}%
where $\gamma$ denotes the Dirac matrices, $\sigma$ the generators of the Clifford basis, and $q_\nu$ the four-momentum transferred in the \tch process.

Each of the operators contributes to the \Wtb\ vertex with a different Lorentz structure: $\mathcal{O}_{\varphi Q}^{3}$ has a Lorentz structure similar to the SM and therefore modifies the left-handed vector coupling ($V_\text{L}$) present in the SM, while $\mathcal{O}_{\varphi tb}$ induces a right-handed vector coupling and $\mathcal{O}_{bW}$ and $\mathcal{O}_{tW}$ contribute as left-handed and right-handed tensor couplings, respectively. Furthermore, $\mathcal{O}_{\varphi tb}$, $\mathcal{O}_{bW}$, and $\mathcal{O}_{tW}$ are non-Hermitian operators and therefore their WCs are complex. A non-vanishing imaginary part then leads to CP violation~\cite{Aguilar_Saavedra_2010}.

The remaining two operators contributing to the top-quark physics studied in this analysis are two four-quark
operators, which modify the top-quark production in the \tch. The relevant four-quark operators have the left--left ($\text{LL}$) and right--right ($\text{RR}$) chiral structure
\begin{align*}
\mathcal{O}_{Qq}^{3,1} = & \left( \bar Q\gamma_{\mu} \tau^I Q \right) \left( \bar q_i \gamma^{\mu} \tau^I q_i \right),\nonumber\\
\mathcal{O}_{Qq}^{3,8} = & \left( \bar Q\gamma_{\mu} T^A \tau^I Q \right) \left( \bar q_i \gamma^{\mu} T^A \tau^I q_i \right),
\end{align*}%

where $Q=(t_\text{L},b_\text{L})$, while $q$ refers to left-handed doublets in the first two generations.
The first operator is an SU(2) triplet and colour singlet and the second one is an SU(2) triplet and colour octet. From these two, only the first operator interferes with the SM amplitude, since \PW-boson exchange is also a colour singlet.
The colour-octet operator has already been constrained quite tightly using processes where it does interfere with the SM amplitude~\cite{TOPQ-2018-11,TOPQ-2020-06,CMS-TOP-19-001,TOPQ-2020-20}.
Therefore, it is assumed that the colour-singlet operator dominates, and this analysis is restricted to studying the operator $\mathcal{O}_{Qq}^{3,1}$.
For convenience, the WC of $\mathcal{O}_{Qq}^{3,1}$, $c_{Qq}^{3,1}$, is referred to as $c_{Qq}$ in this paper.

As seen above, the dimension-six operators contributing to the top-quark interactions considered in this analysis are
\begin{equation*}
\mathcal{O}_{\varphi Q}^{3}, \qquad  \mathcal{O}_{\varphi tb}, \qquad \mathcal{O}_{bW}, \qquad \mathcal{O}_{tW}, \qquad \mathcal{O}_{Qq}^{3,1}.
\end{equation*}
The corresponding EFT coefficients $c_{\varphi Q}$ and $c_{Qq}$ are real, while $c_{\varphi tb}$, $c_{bW}$, and $c_{tW}$ are complex, giving a total of eight free parameters. %
However, the measurement of the angular distributions can only constrain a subspace of the eight-dimensional complex parameter space.
At LO, there are four distinct amplitudes for top-quark decay.
Two of them correspond to decays to a left-handed \bquark and for $m_{\Pqb}=0$,  only receive contributions from the SM, $c_{\varphi Q}$ and $c_{tW}$. The others correspond to decays to a right-handed \bquark and only receive contributions from $c_{\varphi tb}$ and $c_{bW}$.
Thus, one can in principle determine the relative phase between $c_{\varphi tb}$ and $c_{bW}$, but one cannot determine the phase relative to the SM for either.
For convenience, the analysis takes $c_{\varphi tb}$ to be real
and positive, and interprets $\mathrm{Arg}\left(c_{bW}\right)$ as implying a phase relative to that of $c_{\varphi tb}$.
This leaves two imaginary parameters that, in principle, can be constrained by analysing the angular variables. In total, seven free parameters~\cite{Aguilar-Saavedra:2017wpl} remain to be measured: \cpq, \ctW, \IctW, \cbW, \IcbW, \cptb, and \cqq, with:
\footnote{In this work, lowercase $c$ is used for complex‐valued WCs, while uppercase $C$ is reserved for parameters that are strictly real.}
\begin{align*}
V_\text{L} &= V_{tb}\left(1 + \frac{v^2}{\Lambda^2} \cpq\right) \,, \\
c_{tW} &= \ctW + i \IctW \,, \\
c_{bW} &= \cbW + i \IcbW \,, \\
c_{\varphi tb} &= \cptb \,, \\%
c_{Q,q} &= \cqq \,,
\end{align*}

where $V_\text{L}$ is the left-handed vector coupling in the anomalous-coupling notation~\cite{Aguilar-Saavedra:2006qvv} and $V_{tb}$ is the quark-mixing matrix element (ME) in the Cabibbo–Kobayashi–Maskawa (CKM) matrix~\cite{ParticleDataGroup:2024cfk}. The superscript ${}^I$ on the WCs $C$ denotes their imaginary part.

Fixing $c_{\varphi tb}$ to be real and positive is done purely for practical purposes and does not constrain $c_{\varphi tb}$ to have positive or even real values. Multiplying $c_{\varphi tb}$ by an arbitrary phase does not change the posterior probability distributions of the fit as long as $c_{bW}$ is multiplied by the same phase.

Finally, it is noted that $\mathcal{O}_{\varphi Q}^{3}$ has the same Lorentz structure as the SM top-quark decay, and therefore only affects the overall cross-section, and has no effect on the angular distributions.  Thus, measuring normalised angular distributions only constrains the ratios of coefficients $\ctW/V_\text{L}$, $\IctW/V_\text{L}$, etc., while measuring the total rate allows one to set constraints on the absolute values of the coefficients. In this analysis, the strength of measuring both the normalised angular distribution and the cross-section is used to obtain the best sensitivity among all the targeted WCs.

% End of text imported from the .//sections/motivation.tex input file

%
\section{ATLAS detector}
\label{sec:atlas_detector}

% The next lines are included from the .//sections/atlas-detector.tex input file
%
\newcommand{\AtlasCoordFootnote}{%
ATLAS uses a right-handed coordinate system with its origin at the nominal interaction point (IP)
in the centre of the detector and the \(z\)-axis along the beam pipe.
The \(x\)-axis points from the IP to the centre of the LHC ring,
and the \(y\)-axis points upwards.
Polar coordinates \((r,\phi)\) are used in the transverse plane,
\(\phi\) being the azimuthal angle around the \(z\)-axis.
The pseudorapidity is defined in terms of the polar angle \(\theta\) as \(\eta = -\ln \tan(\theta/2)\) and is equal to the rapidity
$ y = (1/2) \ln[(E + p_z)/(E - p_z)] $ in the relativistic limit.
Angular distance is measured in units of \(\Delta R \equiv \sqrt{(\Delta y)^{2} + (\Delta\phi)^{2}}\).}

The ATLAS experiment~\cite{PERF-2007-01} at the LHC is a multipurpose particle detector
with a forward--backward symmetric cylindrical geometry and nearly \(4\pi\) coverage in
solid angle.\footnote{\AtlasCoordFootnote}
It consists of an inner tracking detector surrounded by a thin superconducting solenoid
providing a \qty{2}{\tesla} axial magnetic field, electromagnetic and hadronic calorimeters, and a muon spectrometer.
The inner tracking detector covers the pseudorapidity range \(|\eta| < 2.5\).
It consists of silicon pixel and microstrip detectors, and a transition radiation tracker (TRT).
Lead/liquid-argon (LAr) sampling calorimeters provide electromagnetic (EM) energy measurements
with high granularity within the region \(|\eta|< 3.2\).
A steel/scintillator-tile hadronic calorimeter covers the central pseudorapidity range (\(|\eta| < 1.7\)).
The endcap and forward regions are instrumented with LAr calorimeters
for EM and hadronic energy measurements up to \(|\eta| = 4.9\).
The muon spectrometer surrounds the calorimeters and is based on
three large superconducting air-core toroidal magnets with eight coils each.
The field integral of the toroids ranges between \num{2.0} and \qty{6.0}{\tesla\metre}
across most of the detector.
The muon spectrometer includes a system of precision tracking chambers up to \(|\eta| = 2.7\) and fast detectors for triggering up to \(|\eta| = 2.4\).
The luminosity is measured mainly by the LUCID-2~\cite{LUCID2} detector, which is located close to the beam pipe.
A two-level trigger system is used to select events~\cite{TRIG-2016-01}.
The first-level trigger is implemented in hardware and uses a subset of the detector information
to accept events at a rate below \qty{100}{\kHz}.
This is followed by a software-based trigger that
reduces the accepted event rate to about \qty{1.25}{\kHz} on average
depending on the data-taking conditions.
A software suite~\cite{SOFT-2022-02} is used in data simulation, in the reconstruction
and analysis of real and simulated data, in detector operations, and in the trigger and data acquisition
systems of the experiment.

%
%
%
%
%
%
%
%
%
%
%
%
%
%
%
%
%
%
%
%
%
%
%
%
%
%
%
%
%
%
%
%
%
%
%
%
%
%
%
%
%
%
%
%
%
%
%
%

% End of text imported from the .//sections/atlas-detector.tex input file

%
\section{Data and simulated event samples}
\label{sec:data_mc_samples}

% The next lines are included from the .//sections/data_and_MC_samples.tex input file
%

\newcommand*{\onelep}{\ensuremath{1\Pl}\xspace}
\newcommand*{\dilep}{\ensuremath{2\Pl}\xspace}
\newcommand*{\dilepSS}{\ensuremath{2\Pl\,\text{SS}}\xspace}
\newcommand*{\trilep}{\ensuremath{3\Pl}\xspace}
\newcommand*{\ttWj}{\ensuremath{t\bar{t}W+j}\xspace}

This analysis uses data from \pp collisions at $\sqrt{s} = \SI{13}{\TeV}$ collected with the ATLAS detector
at the LHC during 2015--2018. Stringent detector and data quality requirements were applied~\cite{DAPR-2018-01},
resulting in a data sample corresponding to a total integrated luminosity of \lumi~\cite{LUCID2}.
The events were selected by single-lepton (electron or muon) triggers~\cite{TRIG-2018-05, TRIG-2018-01}, imposing low thresholds on the transverse energy ($\ET$) of electrons and the \pT of muons, in addition to isolation requirements, or imposing a looser identification criterion and higher thresholds with no isolation requirement. The lowest thresholds were varied from 20 to \SI{26}{\GeV}  depending on the lepton flavour and the data-taking period.

Samples of simulated events were produced using different MC event generators, including parton shower (PS) and hadronisation models for both the signal and background processes.
The effect of multiple interactions in the same and neighbouring bunch crossings (pile-up)
was modelled by overlaying the hard-scattering event with simulated inelastic \pp collisions %
generated with \PYTHIA[8.186]~\cite{Sjostrand:2007gs} using the \NNPDF[2.3lo] set of parton distribution functions (PDFs)~\cite{Ball:2012cx} and the A3 set of parameters (tune)~\cite{ATL-PHYS-PUB-2016-017}.
Events were reweighted such that the
distribution of the average number of interactions per bunch crossing
matches that observed in data. \Cref{tab:mcconfig} presents a summary of all configurations used to generate events for signal (\tch) and background processes.
%

% The next lines are included from the .//tables/mc-settings.tex input file
\begin{table}[!htb]
\begin{center}
\caption{The configurations used to generate events for signal
(\tch) and background processes. \enquote{ME order} refers to
the order of the matrix-element calculation. The simulation
samples used to model EFT effects are shown in grey.
A dedicated set of tuned PS parameters~\cite{Bothmann:2019yzt}, developed by
the \SHERPA authors and referred to as the “default” tune, was used
for simulation samples generated with \SHERPA.}
\label{tab:mcconfig}
\vspace{0.25cm}
\small
\setlength\tabcolsep{1.5pt}
\begin{tabular}{~l^l^l^l^l^l}
\toprule
\textbf{Process} & \textbf{Generator} & \textbf{ME order} & \textbf{PS} & \textbf{PDF set} & \textbf{PS tune} \\
\midrule
\tch & \POWHEGBOX~\cite{Frixione:2007nw,Nason:2004rx,Frixione:2007vw,Alioli:2010xd} & NLO & \PYTHIA[8] & \NNPDF[3.0nlo] & A14~\cite{ATL-PHYS-PUB-2014-021} \\
\rowstyle{\color{gray}}
& \MADGRAPH~\cite{Alwall:2014hca} & LO & \PYTHIA[8] & \NNPDF[3.0nlo] & A14    \\
\midrule
\ttbar & \POWHEGBOX & NLO & \PYTHIA[8] & \NNPDF[3.0nlo] & A14  \\
\rowstyle{\color{gray}}
& \MADGRAPH & LO & \PYTHIA[8] & \NNPDF[3.0nlo] & A14  \\
$V+\text{jets}$ & \Sherpa[2.2.11]~\cite{Bothmann:2019yzt} & \MEPSatNLO & \Sherpa & \NNPDF[3.0nnlo] & default \\
Associated $tW$ & \POWHEGBOX & NLO & \PYTHIA[8] & \NNPDF[3.0nlo] & A14   \\
$s$-channel & \POWHEGBOX & NLO & \PYTHIA[8] & \NNPDF[3.0nlo] & A14   \\
$VV$, $qqVV$ & \Sherpa[2.2.2]  & \MEPSatNLO & \Sherpa & \NNPDF[3.0nnlo] & default  \\
$VVV$ & \Sherpa[2.2.2]  & \MEPSatNLO & \Sherpa & \NNPDF[3.0nnlo] & default  \\
\ttH & \POWHEGBOX  & NLO & \PYTHIA[8]\ & \NNPDF[3.0nlo] & A14 \\
$\ttbar W$ & \Sherpa[2.2.10] & NLO QCD+EW & \Sherpa &  \NNPDF[3.0nnlo] & default \\
\ttZ & \MGNLO~\cite{Alwall:2014hca} & NLO & \PYTHIA[8] & \NNPDF[3.0nlo] & A14  \\
$tZ$ & \MGNLO & NLO & \PYTHIA[8] & \NNPDF[3.0nlo] & A14 \\
$tWZ$                  & \MGNLO & NLO DR1 & \PYTHIA[8] & \NNPDF[3.0nlo] & A14 \\
$t\bar t t$      & \MGNLO & QCD NLO & \PYTHIA[8] & \NNPDF[2.3lo] & A14 \\
$t\bar t t\bar t$      & \MGNLO & QCD NLO & \PYTHIA[8] & \NNPDF[3.1nlo] & A14 \\
$\ggH, \qqH$ & \POWHEGBOX & QCD NLO & \PYTHIA[8] & \CT[10]~\cite{Pumplin:2002vw} & \AZNLO~\cite{STDM-2012-23} \\
$\WH, \ZH$ & \PYTHIA[8] & QCD NLO & \PYTHIA[8] & \NNPDF[2.3lo] & A14 \\
\bottomrule
\end{tabular}
\end{center}
\end{table}

% End of text imported from the .//tables/mc-settings.tex input file

The effects of variations associated with different sources of modelling systematic
uncertainty were evaluated by comparing alternative simulation samples with the
nominal ones. These alternative samples were generated either through internal
reweighting or with dedicated event generators. Each variation was introduced
independently unless explicitly stated otherwise, ensuring a consistent
assessment of its individual impact on the analysis. Further details about these samples
are provided in \cref{sec:systematics}.

Single-top-quark production in the \tch was modelled using the
\POWHEGBOX[v2]~\cite{Frederix:2012dh,Nason:2004rx,Frixione:2007vw,Alioli:2010xd}
generator, which provides ME calculations at NLO
accuracy in the strong coupling constant \alphas\ in the 4FS with the corresponding \NNPDF[3.0nlo]~\cite{Ball:2014uwa} set of PDFs.
The functional form of the
renormalisation (\muR) and factorisation (\muF) scales was set to
\(\sqrt{m_b^2 + \pTX[2][b]}\) following the
recommendations of Refs.~\cite{Alioli:2009je, Frederix:2012dh}. Top quarks were decayed at
LO using \MADSPIN~\cite{Frixione:2007zp,Artoisenet:2012st} to preserve
all spin correlations.  For PS and hadronisation, the events were interfaced with
\PYTHIA[8.230]~\cite{Sjostrand:2014zea}, which used the A14
tune~\cite{ATL-PHYS-PUB-2014-021} and the \NNPDF[2.3lo] PDF set~\cite{Ball:2012cx}.

The predicted cross-sections for single-top-quark and single-top-antiquark production in the \tch in \pp collisions at a centre-of-mass energy of 13\,TeV are $\sigma(tq)=134.2^{\,+2.6}_{-1.7}$\,pb and $\sigma(\bar{t}q)=80.0^{\,+1.8}_{-1.4}$\,pb, respectively. The cross-sections were calculated with the MCFM program~\cite{Campbell:2020fhf} at next-to-next-to-leading order (NNLO) in quantum chromodynamics (QCD). %
The quoted uncertainties include the uncertainties due to choosing \muR\ and \muF\ scale values, the uncertainty in the PDFs, and the uncertainty in the value of the strong coupling constant \alphas. The uncertainty due to the scales was determined by raising and lowering \muR and \muF independently by a factor of two, whilst never allowing them to differ by more than a factor of two. The combined PDF and \alphas uncertainties were determined at the 68\% confidence level (CL) according to the Hessian representation of the \PDFforLHC[21] PDF set~\cite{Ball:2022oua}.

The production of \ttbar\ events was modelled at NLO in QCD in the 5FS using
\POWHEGBOX[v2]~\cite{Frixione:2007nw,Nason:2004rx,Frixione:2007vw,Alioli:2010xd} with the \NNPDF[3.0nlo] PDF set.
The \texttt{\POWHEG:hdamp} parameter was set to $1.5 \times \mt$~\cite{ATL-PHYS-PUB-2016-020}.\footnote{The \texttt{\POWHEG:hdamp} parameter
controls the \pT\ of the first additional emission beyond LO
in the PS and therefore regulates the  high-\pT\ emission against which the \ttbar\ system recoils.}
Events were interfaced with \PYTHIA[8.230], which used the A14 tune and the \NNPDF[2.3lo] PDF set~\cite{Ball:2012cx}.
The simulation samples were normalised to the cross-section of $834^{\,+29}_{-37}$~pb predicted at NNLO in QCD, including the resummation of
next-to-next-to-leading logarithmic (NNLL) soft-gluon terms calculated using \TOPpp[2.0]~\cite{Beneke:2011mq,Cacciari:2011hy,Baernreuther:2012ws,Czakon:2012zr,Czakon:2012pz,Czakon:2013goa,Czakon:2011xx}. %
The combined PDF and $\alphas$ uncertainties were determined at the 68\% CL according to the Hessian representation of the \PDFforLHC[21] PDF set.
To model the EFT effects, LO SM events were produced with \MADGRAPH[2.8.1]~\cite{Alwall:2014hca} with the \NNPDF[3.0nlo] PDF set interfaced with \PYTHIA[8.244], which used the A14 tune and the \NNPDF[2.3lo] PDF set.
The events were subsequently reweighted at LO by rescaling each event by the ratio of EFT to SM MEs~\cite{Gainer:2014bta}, computed using the \texttt{dim6top} UFO model~\cite{Aguilar-Saavedra:2018ksv}.
In this setup, the \texttt{dim6top} model is used with EFT insertions allowed at both the production and decay vertex of the top quark.
This setup therefore retains all contributions up to quadratic order in the WCs in the cross-section, arising from the squared ME.
In this approach, events with different parton multiplicities in the \MADGRAPH\ simulation sample are each reweighted using the tree-level ME evaluated at the corresponding final-state multiplicity; this reweighting is applied solely within the LO \MADGRAPH\ simulation sample and does not involve the NLO \POWHEG\ simulation.
Higher-order QCD corrections are accounted for by an additive yield correction defined as the difference between the NLO \POWHEG\ and LO \MADGRAPH\ predictions at the SM point, treated as a constant $k$-factor independent of the WC values.
Consequently, the modelling includes the interference between SM and EFT amplitudes, as well as all interference terms between EFT contributions arising in the production and decay of the top quark.
A six-dimensional
grid, formed by \ctW, \IctW, \cbW, \IcbW, \cptb, and \cqq, was generated for both the \tch signal sample and the \ttbar background sample.
Here, \cpq is not included in the grids, since it only changes the overall cross-section, which can be parameterised by a scale factor, as introduced in~\cref{sec:motivation}.
The points of this grid were designed to cover the $95\%$ confidence intervals of the previous analysis results~\cite{TOPQ-2018-10,TOPQ-2015-11}.
An analytically derived top-quark width correction was incorporated into the reweighting
to account for variations in the top-quark's width $\Gamma_t$ induced by changes
in the WCs~\cite{deBeurs:2630406}, since this correction cannot be handled by the \MADGRAPH[2.8.1] reweighting function.
This procedure ensures a consistent treatment of EFT effects in both production and decay, including the dependence of the top-quark width on the WCs, thereby ensuring that the total width scales correctly with the modified couplings.
The approach follows the same formalism as presented in Ref.~\cite{TOPQ-2018-10}, where the dependence of the cross-section on the WCs is explicitly constructed.
Overall, the signal model consistently includes all SM, linear EFT, and quadratic EFT contributions, together with their mutual interferences, for the full process including both production and decay.

The production of single-boson ($V+\text{jets}$) events was simulated using \SHERPA[2.2.11]~\cite{Bothmann:2019yzt}. The MEs were calculated at NLO (LO) accuracy in QCD for up to two (for three or four) partons (0,1,2j@NLO + 3,4j@LO) using the \textsc{Comix} library~\cite{Gleisberg:2008fv}, with the virtual QCD loop corrections supplied by the \OPENLOOPS library~\cite{Buccioni:2019sur,Cascioli:2011va,Denner:2016kdg}. The ME calculations were matched and merged with the default \SHERPA PS~\cite{Schumann:2007mg}, based on Catani–Seymour dipole factorisation and the cluster hadronisation model~\cite{Winter:2003tt}. The \NNPDF[3.0nnlo] PDF set was used together with a dedicated set of \SHERPA-tuned PS parameters, and the simulation samples were normalised to NNLO cross-section predictions~\cite{Anastasiou:2003ds}.

Single top quarks from $s$-channel and \tW associated production were generated with
\POWHEGBOX[v2] in the 5FS at NLO in QCD with the \NNPDF[3.0nlo] PDF set, with
\muF and \muR set to $\sqrt{m_b^2 + \pTX[2][b]}$, following the
recommendation of Refs.~\cite{Alioli:2009je,Re:2010bp}, respectively. For the latter process,
the Diagram Removal 1~(DR1) scheme was employed to handle the interference between \tW\ and
\ttbar~\cite{ATL-PHYS-PUB-2016-020}.
The EFT effects in the \tW process are neglected, since its impact on the total event yield is small compared to the dominant \tch and \ttbar processes, as demonstrated in the yield tables and figures in the following sections.

The $VV$ and $VVV$ processes were simulated by \SHERPA[2.2.2]~\cite{Bothmann:2019yzt} with off-shell effects and Higgs contributions at NLO (LO accurate for up to three additional parton emissions). The MEs were matched to PS with \MEPSatNLO~\cite{Hoeche:2011fd, Hoeche:2012yf, Catani:2001cc, Hoeche:2009rj} with the set of tuned parameters developed by the \SHERPA authors. The samples include events from loop-induced and EW production.

Top-quark associated and Higgs-boson backgrounds were simulated at NLO in QCD using various generators with matched PS. They include \ttH, \ttW, \ttZ, \tZq, \tWZ, \ttt, \tttt, and Higgs-related processes (\ggH, \qqH, \WH, \ZH).

In all simulated signal and background samples, the top-quark mass was set to
$\mt = \SI{172.5}{\GeV}$. The strong coupling constant was set to $\alphas = 0.118$. These values were also used in the computation of production
cross-sections for processes involving the top quark. The bottom and charm hadron
decays were simulated with \EVTGEN~\cite{Lange:2001uf} for non-\SHERPA-generated
event samples.

%
%
%
%
%

% End of text imported from the .//sections/data_and_MC_samples.tex input file

%
\section{Object definitions}
\label{sec:object}

% The next lines are included from the .//sections/object_definition.tex input file
%

Interaction vertices from the \pp collisions are reconstructed from at least two tracks with $\pt > \SI{500}{\MeV}$
that are consistent with originating from the beam collision region in the $x{-}y$ plane.
If more than one primary vertex candidate is found in the event,
the candidate for which the associated tracks form the largest sum of squared $\pt$ is selected as the hard-scatter primary vertex~\cite{ATL-PHYS-PUB-2015-026}.

Electron candidates are reconstructed from energy clusters in the electromagnetic calorimeter matched to a track in the inner tracking detector (ID)~\cite{EGAM-2018-01}.
They are required to satisfy $\pt>\SI{10}{\GeV}$ and $|\eta_{\mathrm{cluster}}|<2.47$,
excluding the transition region between the barrel and endcap calorimeters ($1.37<|\eta_{\mathrm{cluster}}|<1.52$).
The \texttt{loose} and \texttt{tight} electron identification working points are used to identify Loose and Tight electrons respectively~\cite{EGAM-2018-01}, based on a likelihood discriminant employing calorimeter,
tracking, and combined variables that provide separation between electrons and jets, and electrons and photons.

Muon candidates are reconstructed from tracks measured in the ID matched to tracks measured in the muon spectrometer~\cite{MUON-2018-03,MUON-2022-01}.
The resulting muon candidates are re-fitted using the complete track information from both detector systems.
They are required to satisfy $\pt > \SI{10}{\GeV}$ and $|\eta| < 2.5$.
The \texttt{medium} muon identification working point~\cite{MUON-2022-01} is used.%

Electron (muon) candidates are matched to the primary vertex by requiring that the significance of their transverse impact parameter,
$d_0$, satisfies $|d_0/\sigma(d_0)|<5.0\,(3.0)$, where $\sigma(d_0)$ is the measured uncertainty in $d_0$, and by requiring that their longitudinal impact parameter,
$z_0$, satisfies $|\zzsth|<\SI{0.5}{mm}$~\cite{EGAM-2018-01,MUON-2018-03}.\footnote{The transverse impact parameter, $d_0$, is defined in the $x$--$y$~plane as the distance of closest approach of the track to the beamline. The longitudinal impact parameter, $z_0$, is defined as the distance in $z$ between the primary vertex and the point on the track used to evaluate $d_0$.}
Furthermore, lepton isolation is required for both electron and muon candidates~\cite{EGAM-2019-01,MUON-2018-03}.
Tight leptons with incorrect charge assignment, or originating from heavy-flavour hadron decays, misidentified jets, or photon conversions (collectively referred to as \enquote{non-prompt leptons}) are suppressed using a BDT discriminant, based on calorimeter, tracking, isolation, and lifetime-related quantities~\cite{EGAM-2019-01,MUON-2018-03}, which allows prompt muons (barrel/endcap electrons) to be selected with an efficiency of about 60\% ($60\%/70\%$) at $\pt \sim \SI{20}{\GeV}$ and reaching a plateau of approximately 95\% ($95\%/90\%$) at $\pt \sim 40\ (40/65)~\SI{}{\GeV}$, while achieving a rejection factor of about 20 against leptons from $b$-hadron decays.

\Cref{tab:lepdef} summarises the definitions of the reconstructed electron and muon candidates.

% The next lines are included from the .//tables/lepton_definitions.tex input file
\begin{table}[!htbp]
\begin{center}
\caption{Summary of the electron and muon object definitions. The selection requirements for Tight
electrons/muons are applied in addition to the preselected
(Loose) objects used for the overlap removal.}
\vspace{0.25cm}
\small
\setlength\tabcolsep{1.5pt}
\begin{tabular}{ l|l|l }
\toprule
& \multicolumn{1}{c|}{\textbf{Loose electron}} & \multicolumn{1}{c}{\textbf{Loose muon}} \\
\midrule
Identification   & \multicolumn{1}{c|}{\texttt{loose}} & \multicolumn{1}{c}{\texttt{medium}} \\
Acceptance       & \multicolumn{1}{c|}{$\pt > \SI{10}{\GeV}$, $|\eta^\mathrm{cluster}| < 2.47$}  & \multicolumn{1}{c}{$\pt > \SI{10}{\GeV}$, $\abseta < 2.5$} \\
&  \multicolumn{1}{c|}{except $1.37 < |\eta^\mathrm{cluster}| < 1.52$}          & \multicolumn{1}{c}{}\\
Impact parameter & \multicolumn{1}{c|}{$|\dzero/\sigma(\dzero)| < 5.0$}                   & \multicolumn{1}{c}{$|\dzero/\sigma(\dzero)| < 3.0$} \\
& \multicolumn{1}{c|}{$|\zzsth| < \SI{0.5}{\mm}$}                      & \multicolumn{1}{c}{$|\zzsth| < \SI{0.5}{\mm}$} \\
\midrule
& \multicolumn{1}{c|}{\textbf{Tight electron}} & \multicolumn{1}{c}{\textbf{Tight muon}} \\
\midrule
Identification   & \multicolumn{1}{c|}{\texttt{tight}} &  \multicolumn{1}{c}{\texttt{medium}}\\
Isolation        & \multicolumn{1}{c|}{required}  & \multicolumn{1}{c}{required} \\
Extra selections & \multicolumn{1}{c|}{Electron charge-misassignment veto} &  \multicolumn{1}{c}{} \\
& \multicolumn{1}{c|}{Electron conversion candidate veto} &  \multicolumn{1}{c}{} \\
\bottomrule
\end{tabular}
\label{tab:lepdef}
\end{center}
\end{table}
%

% End of text imported from the .//tables/lepton_definitions.tex input file

%
%
%
%
%
%
%
%
%

%
%
%
%
%
%

%
%
%
%
%
%
%
%
%
%
%
%
%
%
%
%
%
%
%
%
%
%
%
%
%
%
%
%
%
%

The constituents for jet reconstruction are identified by combining measurements from both the ID and the calorimeter using a particle flow (PFlow) algorithm~\cite{PERF-2015-09}.
Jet candidates are reconstructed from these PFlow objects using the \antikt algorithm~\cite{Cacciari:2008gp,Fastjet} with a radius parameter of $R=0.4$.
They are calibrated using simulation with corrections obtained by using in situ techniques in data~\cite{JETM-2018-05}.
Only jet candidates with $\pT> \SI{30}{\GeV}$ and within $|\eta|<4.5$ are selected.
To reduce the effect of \pileup, each jet with $\pT<\SI{60}{\GeV}$ and $|\eta|<2.4$ must satisfy the \texttt{Tight} working point of the jet vertex tagger (JVT)~\cite{PERF-2014-03} criteria.
Moreover, jets with $2.5 < \abseta < 4.5$ (hereafter called \enquote{forward jets}) and $\pT < \SI{120}{\GeV}$ must satisfy the \texttt{Tight} working point of the forward jet vertex tagger (fJVT)~\cite{PERF-2016-06} criteria.
A set of quality criteria is also applied to reject events containing at least one jet arising from non-collision sources or detector noise~\cite{ATLAS-CONF-2015-029}.

Jets containing \Pqb-hadrons are identified (\btagged) via the \texttt{DL1r}~\cite{FTAG-2019-07} algorithm. It uses a deep feed-forward neural network
based on the distinctive features of \Pqb-hadron decays,
primarily the impact parameters of tracks and the displaced vertices reconstructed in the ID.
Additional input to this network is provided by discriminating variables constructed by a recurrent neural network (RNN)~\cite{FTAG-2019-07},
which exploits the spatial and kinematic correlations between tracks originating from the same \Pqb-hadron.
A multivariate \btag discriminant value is calculated for each jet.
In this measurement, a jet is considered \btagged if it passes the working point corresponding to 60\% average expected efficiency to tag a \bquark jet, with a light-flavour-jet rejection factor of about 2500, and a charm-jet rejection factor of about 40.
Correction factors, derived from dedicated calibration samples enriched in \bjet{}s, \Pqc-jets, or light-flavour jets, are applied to the simulated event samples~\cite{FTAG-2018-01,FTAG-2020-08,FTAG-2019-02}.

The reconstruction of hadronically decaying $\tau$-lepton candidates (\tauhad) is seeded by jets reconstructed via the anti-$k_t$ algorithm, using calibrated energy clusters as inputs, with a radius parameter of $R = 0.4$~\cite{ATLAS-CONF-2017-029, ATL-PHYS-PUB-2019-033}.
Jets seeding \tauhad candidates are additionally required to have $\pt > \SI{10}{\GeV}$ and $|\eta|< 2.5$.
To separate the \tauhad candidates initiated by hadronic $\tau$-lepton decays from jets initiated by quarks or gluons, an RNN~\cite{ATL-PHYS-PUB-2019-033}
identification method was trained on information from reconstructed charged-particle tracks and clusters of energy in the calorimeter matched to \tauhad candidates, as well as on high-level discriminating variables.
A separate multivariate discriminant based on a BDT~\cite{ATLAS-CONF-2017-029, ATL-PHYS-PUB-2022-044} is also used to reject backgrounds arising from electrons mimicking a \tauhad.
This discriminant (\texttt{eBDT}) is built using information from the calorimeter and the ID.
Transition radiation information from the TRT system plays a key role in the performance of this discriminant.
Baseline \tauhad are required to have one or three associated tracks, electric charge of $\pm1$, $\pt > \SI{20}{\GeV}$ and $|\eta|< 2.5$, excluding the barrel--endcap transition region.
Each \tauhad must satisfy the \texttt{Medium} RNN identification requirement, with efficiencies of 75\% for 1-prong decays and 60\% for 3-prong decays,
and the \texttt{Loose} \texttt{eBDT} requirement, with an efficiency of 95\%~\cite{ATLAS-CONF-2017-029, ATL-PHYS-PUB-2022-044}.

Ambiguities between independently reconstructed electrons, muons, jets and \tauhad can arise.
A sequential \enquote{overlap removal} procedure is performed to resolve these ambiguities and thus avoid double counting of candidates.
This procedure is applied to leptons satisfying the Loose criteria described in~\cref{tab:lepdef}.
If two electrons are separated by $\Delta R<0.1$, only the one with the higher \pt is kept.
If an electron and a muon overlap within $\Delta R<0.1$, the muon is removed if it is reconstructed only from an ID track and calorimeter energy deposits consistent with a minimum-ionising particle, %
otherwise the electron is removed.
If an electron and a selected jet are found within $\Delta R < 0.2$, the jet is removed.
Any electron subsequently found less than $\DeltaR = 0.4$ from a jet is removed.
A jet--muon separation is required at $\Delta R < 0.4$ if a muon is found within a jet using the ghost-matching technique~\cite{Cacciari:2008gp}.
If the overlapping jet contains less than three tracks with $\pt > \SI{500}{\MeV}$, the jet is removed, otherwise the muon is rejected.
Any \tauhad found less than $\DeltaR = 0.2$ from an electron is removed.
Any \tauhad found less than $\DeltaR = 0.2$ from any type of muon with $\pT > \SI{2}{\GeV}$ is removed,
except those with $\pT > \SI{50}{\GeV}$ which are only removed if overlapping with a combined-type muon.
Finally, if a jet is found less than $\DeltaR = 0.2$ from a \tauhad, the jet is removed.

The missing transverse momentum \ptmiss\ (with magnitude \MET) is defined as the negative vector sum of the transverse momenta of all identified and calibrated objects and remaining unclustered energy,
the latter (composing the \enquote{soft term}) being estimated from low-$p_\text{T}$ tracks associated with the primary vertex but not with any lepton or jet candidate~\cite{JETM-2020-03}.
The \MET is taken as a measurement of the undetectable particles, and is affected by energy losses due to detector inefficiencies and acceptance, and by energy resolution.

% End of text imported from the .//sections/object_definition.tex input file

%
\section{Reconstruction and event selection}
\label{sec:event_selection}

% The next lines are included from the .//sections/reconstruction_and_event_selection.tex input file
The on-shell \PW{} boson originating from the decay of the top quark is
reconstructed from the momenta of the lepton (electron or muon) and the neutrino by
imposing four-momentum conservation. Since the neutrino escapes
undetected, the $x$ and $y$ components of the reconstructed \ptmiss are
assumed to correspond to the \pT\ of the neutrino. The unmeasured
longitudinal component of the neutrino momentum, $p^{\nu}_{z}$, is
computed by imposing a \PW-boson mass constraint\footnote{The constraints used to reconstruct the neutrino momentum are set using the world-average values of the \PW-boson mass, $m_{\PW} = 80.399~\text{GeV}$ from Ref.~\cite{ParticleDataGroup:2010dbb}, which correspond to the value employed in the generation of the MC simulation samples.} on the
lepton--neutrino system. If the resulting quadratic equation yields two real solutions for
$p^{\nu}_{z}$, the solution closer to the reference top-quark mass
$\mt = \SI{172.5}{\GeV}$ is taken. If the
solutions are complex, the reconstructed \ptmiss is rescaled, preserving
its direction, to obtain a unique real solution for
$p^{\nu}_{z}$. The top-quark candidate is then reconstructed by
combining the four-momenta of the reconstructed \PW boson and the
selected \btagged jet.

To enhance the sensitivity to the relevant WCs, it is crucial to maximise the selected yield of \tch single-top-quark events while minimising background contributions from other processes. To achieve this, one signal-enriched region (SR) and two control regions (CRs) are defined. These regions facilitate the simultaneous extraction of the WC values and the data-driven normalisation factors of the \ttbar and \Wjets backgrounds. In the analysis, both the SR and CRs are divided into four channels based on the lepton charge and the \pT of the reconstructed top quark, where channels with reconstructed top-quark \pT greater than \SI{80}{\GeV} are labelled as high-\pT, and those with lower \pT are labelled as low-\pT. Partitioning the data by lepton charge takes advantage of the charge asymmetry present in \Wjets and \tch single-top-quark production, in contrast to the charge-symmetric \ttbar background.
Likewise, separating events by top-quark \pT above and below the \SI{80}{\GeV} threshold, optimised using simulated signal samples, improves the sensitivity because certain EFT operators, most notably \cqQ, impact the high- and low-energy regions differently; for \cqQ, the ratio of the EFT prediction to the SM expectation exhibits a pronounced shape variation, with a sign change around \SI{80}{\GeV}, enhancing the expected constraint by about 20\% in the full likelihood fit.

Events are first required to satisfy a set of preselection criteria, primarily designed to suppress backgrounds from jets misidentified as other objects. The preselection is defined as follows:

\begin{itemize}
\item Exactly one Tight charged lepton $\ell$ (electron or muon, as defined in \cref{tab:lepdef}, originating from the decay of the \PW boson or from the subsequent decay of a leptonically decaying $\tau$) with $\pT > \SI{30}{\GeV}$, matched to the single-lepton trigger.
\item Veto on any additional Loose charged leptons (defined in \cref{tab:lepdef}) to reduce the dilepton backgrounds; events containing reconstructed hadronically decaying $\tau$-leptons (\tauhad) are also vetoed to further suppress processes involving a \tauhad.
\item Exactly two jets; for jets within the calorimeter's endcap--forward transition region ($2.75 < |\eta_j| < 3.5$, where $\eta_j$ denotes the $\eta$ of the jets), the threshold is increased to \SI{35}{\GeV}.%
\item At least one of the jets must be \btagged.%
\item $\MET > \SI{35}{\GeV}$.
\end{itemize}

Two additional multijet background rejection criteria are applied~\cite{TOPQ-2018-10}:

\begin{itemize}
\item Transverse mass of the lepton–\ptmiss system must fulfill $\mtw > \SI{60}{\GeV}$.
\item The lepton transverse momentum $\pT(\ell)$ must satisfy $\pT(\ell) > 50\left[1 - (\pi - |\Delta\phi(j_1, \ell)|)/(\pi - 1)\right]\,\GeV$, where $j_1$ refers to the jet with leading \pT and $\Delta\phi$ is the azimuthal angle between the jet and lepton. This requirement rejects backgrounds, especially the QCD multijet process where two jets
are produced back-to-back in the azimuthal plane and one of those is misreconstructed as a lepton.
\end{itemize}

Events that pass the single-lepton trigger but fail the preselection
are enriched in non-prompt lepton (\enquote{fake}) background contributions and
are therefore used to validate the fake-lepton background estimation.
These backgrounds arise from processes that produce either a jet
misidentified as an electron, or a muon or electron originating
from the semileptonic decay of a heavy-flavour quark. In this
analysis, a matrix method~\cite{EGAM-2019-01} is employed to
estimate such backgrounds in a data-driven way. The parameterised
efficiencies of lepton misidentification are derived from a
multijet-enriched control region requiring a single Loose lepton.
The real- and fake-lepton efficiencies, parameterised in lepton $\pT$ and
$|\eta|$, are then applied event by event to data events with
a Loose, but not Tight, lepton.

Events that satisfy the preselection criteria are classified into three mutually exclusive regions: an SR, a \Wjets-enriched CR (CR-\PW), and a \ttbar-enriched CR (CR-\ttbar). Events in the SR and CR-\PW are required to have exactly one  \btagged jet, with the second jet failing the \btag working point. The CR-\PW region is defined by events that match this \btag configuration but fail at least one of the additional SR selection requirements. Events in CR-\ttbar are required to contain two jets that both satisfy the \btag criteria.

An event is classified as being in the SR if it satisfies the preselection and the following additional requirements:

\begin{itemize}
\item Invariant mass of the lepton and \bjet: $m_{\ell b} <$ \SI{153}{\GeV}; top-quark mass proxy reconstructed from the lepton, missing transverse momentum \ptmiss and \bjet four-vectors: $m_{\Pqt}^{\text{reco}} = m_{\ell\ptmiss b} \in [120.6, 234.6]$~\GeV. These criteria help to select events consistent with a top-quark mass around \SI{172.5}{\GeV}.
\item Trapezoidal requirement defined as: %
\begin{equation*}%
\begin{aligned}
&\left(\eta_j < a_{1} \eta_{\ell\ptmiss b} + b_{1}\right) \cap \left(\eta_j > a_{1} \eta_{\ell\ptmiss b} - b_{1}\right) \cap \\
&\left[ \left(\eta_j > a_{2} \eta_{\ell\ptmiss b} + b_{2}\right) \cup \left(\eta_j < a_{2} \eta_{\ell\ptmiss b} - b_{2}\right) \right],
\end{aligned}
\end{equation*}
where the parameters $a_{1,2}$ and $b_{1,2}$ are optimised by applying the $m_{\ell b}$ requirement and finding the maximum signal-to-background ratio, \soverb. The chosen values are $a_{1} = 3$, $b_{1} = 10.5$, $a_{2} = 0.25$, and $b_{2} = 2.5$. This requirement helps to distinguish the signal process from the \Wjets background.
\item Invariant mass of the full system: $m_{j\ell\ptmiss b} > \SI{320}{\GeV}$.  In \tch single-top-quark events, the angles between these objects are typically large, leading to a high invariant mass, whereas the backgrounds peak at a lower mass value. %
\item Scalar sum of the \pT of all the reconstructed objects in the final state (lepton, jets and \ptmiss): $\HT > \SI{190}{\GeV}$. This ensures the presence of significant hadronic activity.
\item To reject events with semileptonic decayed \bjets, the events are selected by performing a kinematic fit with the reconstructed masses of the \PW boson and the top quark constrained to their measured values~\cite{ParticleDataGroup:2024cfk} in the kinematic likelihood (KL) fit framework~\cite{Erdmann_2014}.%
\footnote{Breit--Wigner distributions are used to construct the kinematic likelihood function. For the \PW boson, the Breit--Wigner distribution of the \lnu system has a mean value of $\mW = \SI{80.399}{\GeV}$~\cite{ParticleDataGroup:2010dbb} and a width of  $\Gamma_{\PW} = \SI{2.1}{\GeV}$. For the top quark, the Breit--Wigner distribution of the \PW\Pqb system has a mean value of $\mt = \SI{172.5}{\GeV}$  and a width of $\Gamma_{t}  = \SI{1.5}{\GeV}$.}
\end{itemize}

\Cref{table:yields_sr_pT_separat} summarises the yields for each channel in the SR, detailing contributions from different MC process components as well as the data.

% The next lines are included from the .//tables/event_yields_pT_separat.tex input file
%
%
%
%
%
%
%
%
%
%
%
%
%
%
%
%
%
%
%
%
%
%
%
%
%
%
%
%
%
%
%
%
%
%
%

\begin{table}[htbp]
\centering
{
\tablesetup
\caption{The pre-fit event yields in the four SRs: \Pqt low-\pt, \Pqt high-\pt, \Paqt low-\pt, and \Paqt high-\pt.
The event yield predictions are obtained from simulated samples normalised to theoretical cross-sections, except for the multijet contribution which is derived from a data-driven method.
Uncertainties quoted include both statistical and systematic components.
Individual predictions are rounded to the relevant digit with the `Total expected' obtained from the sum of the full-precision individual predictions and subsequently rounded.
The last two rows show the signal-to-background (\soverb) and the
data-to-prediction ratios in each region, computed using
full-precision individual predictions.}
\label{table:yields_sr_pT_separat}
\begin{tabular}{l|
*{4}{r@{\,\( \pm \)\,}l}}
\toprule
\textbf{Process} & \multicolumn{2}{c}{\textbf{SR \(t\) low-\pt}} & \multicolumn{2}{c}{\textbf{SR \(t\) high-\pt}} & \multicolumn{2}{c}{\textbf{SR \(\bar{t}\) low-\pt}} & \multicolumn{2}{c}{\textbf{SR \(\bar{t}\) high-\pt}} \\
\midrule
\tch        & \thousand{23900} & 1400 & \thousand{15400} & 1300  & \thousand{13900} & 800 & 7260 & 600 \\
$s$-channel  & 80 & 7 & 20 & 2 & 80 & 5 & 30 & 2 \\
$tW$         & 570 & 20 & 210 & 10 & 570 & 20 & 198 & 10 \\
\ttbar      & 3350 & 450 & 1730 & 240  & 3380 & 540 & 1780 & 230 \\
\Wjets      & 6500 & 700 & 1680 & 200  & 5300 & 800 & 1150 & 150 \\
\Zjets, $VV$ & 600 & 130 & 170 & 32 & 550 & 120 & 160 & 35 \\
Others      & 300 & 110 & 80 & 40  & 400 & 160 & 110 & 40 \\
\midrule
\textbf{Total expected} &  \thousand{35300} & 1600 & \thousand{19300} & 1300 & \thousand{24100} & 1200 & \thousand{10690} & 700 \\
\midrule
\textbf{Data} & \multicolumn{2}{c}{\thousand{35817}}  & \multicolumn{2}{c}{\thousand{18573}}  & \multicolumn{2}{c}{\thousand{24639}} & \multicolumn{2}{c}{\thousand{10536}} \\
\midrule
\textbf{\soverb} & 2.10 & 0.20 & 3.95 & 0.50  & 1.36 & 0.15 & 2.12 & 0.25 \\
\textbf{Data\;/\;Prediction} & 1.02 & 0.05 & 0.96 & 0.07 & 1.02 & 0.05 & 0.99 & 0.06 \\
\bottomrule
\end{tabular}
}
\end{table}

% End of text imported from the .//tables/event_yields_pT_separat.tex input file

%

%

%

%

%

%

%

%
%
%
%
%
%
%
%
%
%
%
%
%
%
%
%
%
%
%
%
%
%
%
%
%
%
%
%
%
%
%
%
%

% End of text imported from the .//sections/reconstruction_and_event_selection.tex input file

%
%
%
%
%
\section{Sources of systematic uncertainty}
\label{sec:systematics}

% The next lines are included from the .//sections/systematics.tex input file
%
%

Various sources of systematic uncertainty are evaluated for their effect on the signal and background rates and the shape of the angular distributions.
In the following, the procedures used to evaluate these sources of systematic uncertainty are discussed. The systematic uncertainties considered are of two main types: experimental uncertainties and theory modelling uncertainties.
All systematic uncertainties are modelled uniformly across the full set of simulation and reconstruction procedures for ATLAS data-taking during the LHC Run 2 scenario.
The effect due to the limited size of the simulated event samples is also taken
into account when evaluating the total uncertainty.
The uncertainty in the integrated luminosity is $\pm0.83$\%~\cite{DAPR-2021-01} and is applied to the signal and all simulated background processes that are scaled by the luminosity as one overall rate uncertainty in the fit.
To account for the difference between the pile-up distributions in data
and MC simulations, an uncertainty related to the scale factor used
to adjust the MC pile-up to the data pile-up profile is applied. This uncertainty is obtained by scaling the value of average number of interactions per bunch-crossing \avgmu in data by factors of $1.01$ and $0.93$ instead of the nominal scale factor of $0.97$~\cite{STDM-2015-05}.%

To determine the jet energy scale (JES) and its associated uncertainty, information from test-beam data, LHC collision data and simulation are used as described in Ref.~\cite{JETM-2018-05}.
The JES uncertainty is decomposed into a set of 31 uncorrelated
components, with contributions from pile-up, jet
flavour composition, single-particle response, effects of jets not
contained within the calorimeter, theory modelling, and statistical uncertainties.
The jet energy resolution (JER) is measured separately for
data and MC simulation, using in situ techniques~\cite{JETM-2018-05}.
Its uncertainty is represented by 13 components
accounting for jet-$\pT$- and jet-$\eta$-dependent differences between
simulation and data.
The uncertainty associated with the
JVT is obtained by increasing and decreasing the scale factor used to correct the JVT
efficiency in simulation within its
uncertainties~\cite{PERF-2016-06}.
The uncertainties associated with the \fJVT are related to data--MC efficiency scale factors in the same manner as the JVT uncertainty.

The efficiency of the \btag algorithm is measured for each jet flavour using control samples in data and in simulation.
From these measurements, correction factors are derived to correct the tagging rates in the simulation.
For \bjet{}s, the correction factors and their uncertainties are estimated from data using dileptonic \ttbar events~\cite{FTAG-2018-01,FTAG-2019-07}.
For \cjets, they are derived from jets arising from \Wboson boson decays in \ttbar events~\cite{FTAG-2020-08}.
For light-flavour jets, the correction factors are derived from \Zjets events~\cite{FTAG-2019-02}.
Uncertainties from sources affecting the \Pqb-tagging efficiencies of \bjets, \cjets and light-jets are evaluated as a function of jet \pt, including bin-to-bin correlations.

For electrons and muons, the reconstruction, identification, isolation, and trigger performance
can differ slightly between data and MC simulation. Scale factors, estimated using the tag-and-probe
method~\cite{EGAM-2018-01,MUON-2018-03,MUON-2022-01}, are applied to simulated events to correct for these differences.
The lepton momentum scales and resolutions are assessed using leptonic decays of $Z$ bosons and \Jpsi mesons.
Corrections to the lepton momentum scale and resolution are
applied to data and simulation respectively, and include
uncertainties in the lepton trigger efficiencies.
The associated systematic uncertainties are then propagated to the
distributions used in this analysis.
Since events containing reconstructed hadronically decaying $\tau$-leptons are vetoed, a \tauhad energy scale systematic uncertainty~\cite{ATLAS-CONF-2017-029} is also included.

The systematic uncertainties related to the modelling of the \MET\ in
the simulation are estimated by propagating the uncertainties in the
energy and momentum scales of electrons, muons and jets, as well as
the uncertainties in the resolution and scale of the soft term~\cite{PERF-2016-07}.
For the data-driven fake-lepton background estimate, %
real-lepton and fake-lepton efficiency differences between the MC-driven and data-driven
methods are taken as an additional source of systematic uncertainty.\footnote{The MC-driven method employs the same matrix method as the data-driven method but replaces the data with simulated samples and recomputes the fake-lepton efficiency. The real-lepton efficiency remains the same for both methods.}
Systematic uncertainties associated with the signal and background MC
modelling are estimated by comparing event samples from different
generators and by varying parameter values in the event generation.

The uncertainty due to the choice of hadronisation model and the other
non-perturbative aspects of the PS is evaluated by comparing samples generated with different hadronisation models or set-ups. The
uncertainty in the PS is estimated for all top-quark processes by comparing
samples from \POWHEGBOX interfaced to \PYTHIA[8] with samples from \POWHEGBOX
interfaced to \HERWIG[7.1.3]. %
To estimate the matching scheme uncertainties in \tch single-top-quark and \ttbar production, a
sample was produced with \POWHEGBOX[v2] and \PYTHIA using a
\texttt{POWHEG:pThard} setting of 1, instead of 0 as in the nominal sample.
This setting regulates the definition of the vetoed region of the \PYTHIA[8]
showering~\cite{Hoche:2021mkv}. The effect of choosing a different
\texttt{\POWHEG:hdamp} parameter value in \ttbar simulation is assessed with
alternative samples generated with $\hdamp = 3.0 \times \mt$.
To assess the uncertainty arising from the choice of recoil scheme in the modelling of QCD radiation from $b$-quarks in top-quark decays, an alternative \ttbar sample is generated with \POWHEGBOX[v2] interfaced to \PYTHIA, in which the \texttt{POWHEG:recoilToTop} option replaces the nominal \texttt{POWHEG:recoilToColor} configuration used for the baseline \ttbar simulation setup~\cite{Brooks:2019xso}.
The effect of the top-quark mass uncertainty is estimated as the difference
resulting from samples with a \SI{0.5}{\GeV} top-quark mass variation, for \tch and \ttbar productions.
All the other modelling uncertainties are
uncorrelated between the \tch\ and \ttbar\ processes, except for the top-quark mass uncertainty.
For associated $\tW$ production, an additional sample using the
diagram subtraction (DS) scheme was produced to evaluate the \ttbar\ interference uncertainties.

The uncertainty due to missing higher-order QCD corrections in the ME
computation is estimated for all processes involving top quarks by independently
varying the \muR\ and \muF\ scales by factors of 0.5 and 2.0 with respect to their
chosen value.
Additionally, for the \tch\ and \ttbar\ processes, uncertainties due to initial-state radiation modelled by the PS are assessed by
varying the \texttt{Var3c} eigentune variation of the A14 tune \cite{ATL-PHYS-PUB-2017-007}.
Furthermore, the impact of final-state radiation is evaluated by varying the \muR\ scale for PS emissions by factors of 0.5 and 2.0, using reweighted events.
For all processes, PDF uncertainties are evaluated using the PDF4LHC15 uncertainty set which
consists of 30 eigenvector variations from multiple NLO PDF sets~\cite{Butterworth:2015oua}.

The event yields associated with the simulated signal and background
processes are estimated using the selection acceptances and the
theoretically predicted cross-sections reported
in \Section{sec:data_mc_samples}. The uncertainties in these
cross-sections are taken into account. However, for those processes
whose normalisation is extracted from the data-driven fits, the effect is negligible.
To evaluate the impact of the flavour composition uncertainty
in the \Wjets simulation templates, an upward and downward variation of
10\% is applied to the relative rate of \PW+\,\bjets vs \PW+\,\cjets.
No variation is considered for
\PW+\,light-flavour~jets production since its contribution is negligible. An uncertainty of $20\%$ is assigned to the \Zjets cross-section. For \ttbar events, an additional uncertainty in the modelling of the top-quark \pT distribution is evaluated by taking the full difference between applying and not applying the correction needed to match the predictions accurate to NNLO in QCD and NLO in EW theory~\cite{Czakon:2017wor}.

Statistical fluctuations in the MC-simulated event samples contribute to the overall systematic
uncertainty. These uncertainties arise from the finite number of
simulated background and signal events.
%
%
%
%
%
%
%
%
%
%
%
%
%
%
%
%
%
%
%
%
%
%
%
%
%
%
%
%
%
%
%
%
%
%
%
%

%
%

%
%
%
%
%

%

%
%

%
%
%
%
%

%

%
%
%
%

%

%

%
%
%
%
%
%
%
%
%
%

%

%
%

% End of text imported from the .//sections/systematics.tex input file

%
\section{EFT analysis strategy}
\label{sec:analysis}

% The next lines are included from the .//sections/analysis.tex input file
%
\def\mhyphen{{\hbox{-}}}

A standard method for modelling EFT effects in differential distributions involves
the use of templates; however, this becomes impractical for a fourfold distribution.
The method used here instead performs a Fourier analysis of the quadruple-differential angular
decay rate of \tch single-top-quark production at detector level, using a set of orthogonal basis functions.
The resulting Fourier coefficients provide a complete characterisation of the
shape of the quadruple-differential angular decay distribution.
They are computed through a simple averaging procedure,
which simultaneously provides their uncertainties and correlations.
Along with the event yields, these coefficients, their uncertainties, and their correlations
serve as inputs to a $\chi^2$ fit to the WCs. An advantage of this approach lies in its ability to
simultaneously capture many features of the shape and normalisation of the
differential cross-section, including those that are more frequently studied
in dedicated analyses, such as cross-section or polarisation analyses.

This analysis procedure is described in more detail in the following sections.
\cref{sec:mfunctions} explains the procedure used to transform the quadruple-differential
angular decay rates into a limited number of coefficients for an orthogonal basis of
so-called \mfunc. \cref{sec:morphing}
describes how to obtain the EFT dependency of the \mfunc coefficients. \cref{sec:fitting}
introduces the statistical tool used to obtain the final limits.

\subsection{Choice of basis functions}
\label{sec:mfunctions}

The orthogonal functions employed in this analysis are called {\mfunc}s
and are defined as
\begin{equation}
M^{j_1 j_2}_{m' m} (\phi,\theta,\phi^*,\theta^*)= \frac{1}{4\pi} ( 2 j_1+1)^{1/2} (2 j_2+1)^{1/2} D_{m'm}^{j_1} (\phi,\theta,0) D_{m,0}^{j_2}(\phi^*,\theta^*,0) \,
\label{ec:M}
\end{equation}
where $ D_{m'm}^{j_1}$ and $D_{m,0}^{j_2}$ are the Wigner $D$-functions~\cite{Martin_1960}.
\begin{equation*}
D^j_{m'm}(\alpha,\beta,\gamma) \equiv \langle jm' | e^{-i \alpha J_z} e^{-i \beta J_y} e^{-i \gamma J_z} | jm \rangle \,.
\end{equation*}
The \mfunc obeys the  orthonormality condition
\begin{equation*}
\iint  M^{j_1 j_2}_{m_1m_2}(\phi,\theta,\phi^*,\theta^*)^* M^{j'_1 j'_2}_{m_1'm_2'}(\phi,\theta,\phi^*,\theta^*)\dif\Omega \dif\Omega^* =\delta_{j_1,j_1'}\delta_{j_2,j_2'}\delta_{m_1,m_1'}\delta_{m_2,m_2'}\,.
\end{equation*}

This choice of basis functions is motivated by the form of the quadruple-differential top-quark decay distribution~\cite{Aguilar-Saavedra:2011jfc}.  Using the helicity formalism of Ref.~\cite{Jacob:1959at}, the amplitude for the top-quark decay $t \to W b \to \ell \nu b$ can be written, in the narrow-width approximation, as
\begin{equation}
A_{\mu \lambda_2 \lambda_3 \lambda_4} = \sum_{\lambda_1} a_{\lambda_1 \lambda_2} b_{\lambda_3 \lambda_4} D_{\mu \nu}^{1/2\,*}(\phi,\theta,0) \, D_{\lambda_1 \lambda}^{1\,*}(\phi^*,\theta^*,0) \,,
\label{ec:A}
\end{equation}
where $\lambda_1$, $\lambda_2$, $\lambda_3$ and $\lambda_4$ are defined to be the helicities of the $W$ boson, $b$-quark, charged lepton and neutrino, respectively, $\mu$ is the third spin component of the top quark, $\nu = \lambda_1 - \lambda_2$, $\lambda = \lambda_3 - \lambda_4$, %
and $a_{\lambda_1 \lambda_2}$ and $b_{\lambda_3 \lambda_4}$ are constants. For the top-quark decay, angular momentum conservation implies that there are only four non-zero reduced amplitudes (or just \enquote{amplitudes} for short):\ $a_{1, \oh}$, $a_{0, \oh}$, $a_{0, \moh}$ and $a_{-1, \moh}$. The reduced amplitudes are simply complex constants unlike the $A_{{\mu \lambda_2 \lambda_3 \lambda_4}}$ of Eq.~(\ref{ec:A}),  which carry the angular dependence. %
The SM interaction of the $W$ boson with charged leptons implies  $\lambda_3 = \pm 1/2$ and $\lambda_4 = \mp 1/2$ for $W^\pm$ decays, assuming massless charged leptons. Therefore, $\lambda = 1$ for top quarks, $\lambda = -1$ for top anti-quarks, and in each case there is only one non-zero $b$ constant, which can be factored out and plays no further role in the distributions.  With this choice of basis functions, only 23 coefficients fully describe the top-quark decay.

Thus, the differential cross-section at parton level
can be fully determined by 23 coefficients
$c^{j_1 j_2}_{m'm}$ defined in Eq.~(\ref{eq:exp}), having  $j_1\le1$ and $j_2\le2$~\cite{Aguilar-Saavedra:2011jfc}: %

\begin{equation}
\frac{1}{\Gamma} \frac{\dif\Gamma}{\dif\Omega \dif\Omega^*} = \sum_{j_1 j_2 m' m} c^{j_1 j_2}_{m'm} M^{j_1 j_2}_{m'm} (\phi,\theta,\phi^*,\theta^*)\,,
\label{eq:exp}
\end{equation}

where $\Gamma$ is the decay rate.

Through a Fourier transform, the $M$-function coefficients can be determined by projecting the differential distribution
$\rho(\phi,\theta,\phi^*,\theta^*) \equiv (1/\Gamma) \dif\Gamma/\dif\Omega \dif\Omega^*$ onto the \mfunc basis:
\begin{equation*}
c^{j_1 j_2}_{m'm} = \int \dif\Omega \dif\Omega^* \rho(\phi,\theta,\phi^*,\theta^*) M^{j_1 j_2}_{m'm}(\phi,\theta,\phi^*,\theta^*)^* \,.
\end{equation*}

The right-hand side of this equation is not only interpreted as a coefficient, but also represents the average of the quantity $M^{j_1 j_2}_{m'm}(\phi,\theta,\phi^*,\theta^*)^*$ over a dataset distributed according to $\rho(\phi,\theta,\phi^*,\theta^*)$. Therefore, the practical approach to determining $c^{j_1 j_2}_{m'm}$ for any given dataset is to first substitute $(\phi_i,\theta_i,\phi^*_i,\theta^*_i)$ into Eq.~(\ref{ec:M}) for each individual event $i$, and then take the average over the entire dataset. This averaging procedure serves as the foundation of this analysis. Hence, the procedure used to obtain these coefficients for both data and MC simulation is to calculate
\begin{equation}
c_{m^\prime m}^{j_1 j_2} = \frac{\sum_{i=1}^N w_i M_{m^\prime m}^{j_1 j_2}\left(\phi_i,\theta_i,\phi^*_i,\theta^*_i\right)}{\sum_{i=1}^N w_i}\,,
\label{eq:mfcomp}
\end{equation}
where the sum is over the event sample, and $w_i$ is the MC weight of the $i^\text{th}$ event, and for data events $w_i = 1$.  The weighted averages yield 23 mean values designated as $\vec{c}^{(s)}$ of channel $(s)$ for real and, when relevant, imaginary parts of \mfunc values over the dataset.

The averaging procedure used in Eq.~\ref{eq:mfcomp} is also applied to products
of real and imaginary parts of $M$-functions to obtain the average of their products over the dataset. Subtracting the product of the averages from the average of the products yeilds the sample covariance, then dividing by the sum of weights (or number of events, in data), gives the covariance of the two observables, or the variance if the two observables are identical. These are elements of the statistical covariance matrix $\mathbb{C}^{(s)}$ which is used below in the construction of the likelihood function.

Detector effects, including resolution, acceptance, and efficiency, strongly affect the angular distributions and hence the $M$-function coefficients. As well as modifying the 23 \mfunc coefficients, these effects
introduce coefficients with $j_1>1$ and/or $j_2>2$\footnote{Angular momentum conservation ensures that, at the generator level, the $\mfunc$ coefficients vanish whenever $j_1 > 1$ and $j_2 > 2$.}.
This analysis is restricted to considering
angular components with $j_1\le1$ and $j_2\le2$, at reconstruction level, since the coefficients with $j_1>1$ and/or $j_2>2$ are mainly shaped by the detector effects, whereas the chosen \mfunc coefficients are mainly
shaped by the EFT operators.
The dependence of these \mfunc coefficients on the WCs is computed at reconstruction level using \MADGRAPHPYTHIA{} MC events as described in~\cref{sec:data_mc_samples}.

Both the signal and the \ttbar background are modelled using MC simulation, and the corresponding \mfunc coefficients are obtained after the background is suppressed by the event selection defined in~\cref{sec:event_selection}.
The background coefficients are then added linearly to the signal \mfunc coefficients, since Fourier analysis is a linear operation, to construct a model for the data. The signal component of the model, as well as the \ttbar background component, are allowed to vary with the WCs, as described in~\cref{sec:morphing}.

\subsection{Morphing}
\label{sec:morphing}

To model the response of the \mfunc coefficients to variations of the WCs, a morphing technique is used. %
The numerator of the right-hand expression in Eq.~(\ref{eq:mfcomp}) is taken to give the unnormalised \mfunc coefficients.
They are proportional to the cross-section of the process, and are each a polynomial function of the WCs.
This parameterisation is mathematically equivalent to the explicit morphing functions used in previous measurements with a smaller number of operators (e.g. Ref.~\cite{TOPQ-2018-10}), but is generalised here to simultaneously accommodate the larger set of WCs considered.
For the WCs \ctw, \citw, \cbw, \cibw and \cptb, the corresponding operators contribute to both the \tch production of a single top quark and its decay.
Therefore, the observables that are proportional to the cross-section depend on these WCs to at most fourth-order, as the amplitude is at most quadratic in the WCs and the cross-section is obtained from its square.

On the other hand, their dependence on \cqq, which only affects the production of single-top-quark events, will be at most second-order.

The unnormalised \mfunc coefficients, for different EFT assumptions, are computed from reconstructed LO \MADGRAPHPYTHIA{} events.
For \cqq, a three-point Lagrange interpolation is used to parameterise the \mfunc; for the other WCs, a five-point Lagrange interpolation is employed.
The correlations among all the WCs are incorporated in the results via the interference terms when using this procedure.
Consequently, the final morphing function for each \mfunc coefficient is expressed as a 203-term polynomial in \ctw, \citw, \cbw, \cibw, \cptb, and \cqq.
This is of the same form as the simpler 15-term polynomial used in Ref.~\cite{TOPQ-2018-10}.

An overall LO-to-NLO correction term for the \mfunc coefficients is calculated by determining the difference between the LO \MADGRAPHPYTHIA{} SM event sample and the NLO \POWBOXPYTHIA{} SM event sample:

\begin{equation*}
\vec{c}_O^{\text{NLO}}(\vec{C}) =   \vec{c}_O^{\text{LO}}(\vec{C}) + \left[\vec{c}_O^{\text{NLO}}(\vec{C}_{\text{SM}}) - \vec{c}_O^{\text{LO}}(\vec{C}_{\text{SM}})\right].
\end{equation*}

The LO-to-NLO event-count correction is obtained as a scale factor equal to the ratio of the two event counts. %

Finally, \cpq is not morphed; instead, the dependency of \cpq could be obtained from the \tch scale factor $S_\text{\tch}$ as described in Eq.~(\ref{eq:cphiq_sftchan}):

\begin{equation}
S_\text{\tch}=|V_\text{L}|^2= \left(1+\frac{v^2}{\Lambda^2} \cpq\right)^2.
\label{eq:cphiq_sftchan}
\end{equation}

\subsection{Fit strategy}
\label{sec:fitting}

Data used in the statistical analysis consist of 23 \mfunc coefficients $\vec{c}^{(s)}$ for each of the four channels $s=1,...,4$, distinguished by top-quark \pT and lepton charge as defined in~\cref{sec:event_selection}, which results in a total of 92 coefficients. In addition, the event counts in each channel's SR, \Wjets CR and \ttbar CR are used, which gives 12 additional inputs.  A statistical covariance matrix in the coefficients is determined by the \mfunc analysis along with their values as described in~\cref{sec:mfunctions}. Because of the manner in which the {\mfunc}s are obtained, i.e.\ Eq.~(\ref{ec:M}), the central limit theorem guarantees Gaussian uncertainties, and a simple $\chi^2$ fit is used. Parameters of the fit include the WCs  \ctw, \citw, \cptb, \cbw, \cibw, and \cqq, collectively referred to as the parameter of interest (POI) vector $\vec{C}$, as well as normalisation factors for the \tch single-top-quark process (namely \cpq as defined in Eq.~(\ref{eq:cphiq_sftchan})) and the \Wjets and \ttbar processes, collectively denoted by the vector $\vec{S}$.

The systematic uncertainty is taken into account by computing a covariance matrix for the systematic uncertainty sources:%
\begin{equation}
\mathbb{C}^\text{syst.} = \sum _\text{syst. sources}
\Delta\vec{v}\otimes\Delta\vec{v}^T,%
\label{eq:systcov}
\end{equation}
where $\Delta\vec{v}$ is the vector of the expected $1\sigma$ uncertainties from the sources of systematic uncertainty described in~\cref{sec:systematics}.
Due to the central limit theorem, the likelihood function combining the \mfunc coefficients and the event counts is a multivariable Gaussian function. The $\chi^2=-2\ln{\mathcal{L}}$ is given by

\begin{equation}
\begin{aligned}
\chi^2(\vec{C},\vec{S})
&= \sum_{(s)} \Biggl\{
\left(\vec{c}^{(s)} - \vec{c}^{(s),\mathrm{model}}(\vec{C},\vec{S})\right)^{T}
\left(\mathbb{C}^{(s)}\right)^{-1}
\left(\vec{c}^{(s)} - \vec{c}^{(s),\mathrm{model}}(\vec{C},\vec{S})\right)
\\
&\qquad\quad
+ \sum_{j \in \{\mathrm{SR},\, \mathrm{CR\text{-}W},\, \mathrm{CR\text{-}t\bar{t}}\}}
\frac{\left(N^{(s,j),\mathrm{model}}(\vec{C},\vec{S}) - N^{(s,j)}\right)^2}{N^{(s,j)}}
\Biggr\},
\end{aligned}
\label{eq:simple_likelihood}
\end{equation}

where $\vec{c}^{(s)}$ is the vector of 23 \mfunc coefficients in channel ${s}$,  $N^{(s,j)}$ is the event count in channel $s$ and SR or CR $j$, and the covariance matrix $\mathbb{C}^{(s)}$ is a
linear sum of statistical covariance $\mathbb{C}_\text{data}^\text{stat.}$ from the data, statistical covariance $\mathbb{C}_\text{MC}^\text{stat.}$ from the MC events, and systematic covariance
$\mathbb{C}^\text{syst.}$ as defined in Eq.~(\ref{eq:systcov}). The superscript \enquote{model} indicates values predicted from the fitting model. The fit has $92+12\space\text{ inputs} - 9\space\text{ parameters} = 95\space\text{ degrees of freedom}$. Systematic uncertainties can be treated either individually or collectively by computing the systematic covariance matrix of Eq.~\ref{eq:systcov} from a restricted set of sources of uncertainty, and introducing the resulting matrix into the $\chi^2$ of Eq.~\ref{eq:simple_likelihood}. The fit is tested with pseudoexperiments over a wide range of input physics and nuisance parameters consistent with the expected precision of the measurement.

%

% The next lines are included from the .//tables/ranking.tex input file

\begin{table}[!htb]
\centering
\caption{Summary of the expected systematic and statistical uncertainty breakdown for the $1\sigma$ confidence interval of each
WC from the one-dimensional likelihood scans. The
limited data and MC simulation sample sizes are sources of statistical
uncertainty. The symbol `$-$' indicates that the uncertainty is negligible.
\label{table:uncertainty_summary}}
\resizebox{0.98\textwidth}{!}{%
\begin{tabular}{ l c c c c c c c c }
\toprule
\textbf{Source of uncertainty} & \textbf{\ctw} & \textbf{\citw} & \textbf{\cptb} & \textbf{\cbw} & \textbf{\cibw} & \textbf{\cqq} & \textbf{\cphiq} \\
\midrule
Jet energy scale & $-0.03, +0.03$ & $-0.01, +0.01$ & $-0.6, +0.4$ & $-0.3, +0.3$ & $-0.3, +0.3$ & $-0.02, +0.05$ & $-0.10, +0.09$ \\
Pile-up & $-0.01, -$ & $-$ & $-0.1, +0.2$ & $-, +0.1$ & $-0.1, -$ & $-, +0.01$ & $-0.08, +0.06$ \\
\btag & $-0.02, +0.03$ & $-0.01, +0.01$ & $-0.6, +0.6$ & $-0.1, +0.2$ & $-0.1, +0.1$ & $-0.02, +0.02$ & $-0.13, +0.13$ \\
Jet energy resolution & $-0.02, +0.02$ & $-0.01, +0.01$ & $-0.8, +0.7$ & $-0.1, +0.1$ & $-0.1, +0.1$ & $-0.02, +0.03$ & $-0.08, +0.09$ \\
PDF & $-0.02, +0.02$ & $-0.01, +0.01$ & $-0.7, +0.6$ & $-0.1, +0.1$ & $-0.1, +0.1$ & $-0.03, +0.04$ & $-0.08, +0.09$ \\
Muon & $-0.03, +0.02$ & $-$ & $-0.4, +0.3$ & $-0.1, +0.1$ & $-0.1, +0.1$ & $-0.01, +0.03$ & $-0.07, +0.06$ \\
\met & $-0.02, +0.02$ & $-$ & $-0.5, +0.4$ & $-0.1, +0.1$ & $-0.2, +0.2$ & $-0.01, +0.02$ & $-0.06, +0.05$ \\
$e/\gamma$ & $-0.01, +0.01$ & $-$ & $-0.2, +0.1$ & $-0.1, +0.1$ & $-0.1, +0.1$ & $-, +0.01$ & $-0.06, +0.07$ \\
Jet vertex tagging & $-0.01, +0.01$ & $-$ & $-0.1, +0.2$ & $-, +0.1$ & $-$ & $-0.02, +0.03$ & $-0.17, +0.18$ \\
Lumi.\ \& cross-section & $-0.01, -$ & $-$ & $-0.1, +0.1$ & $-0.1, +0.1$ & $-0.1, +0.1$ & $-0.01, -$ & $-0.07, +0.07$ \\
\tauhad energy scale & $-$ & $-$ & $-$ & $-$ & $-$ & $-$ & $-$ \\
Non-prompt leptons & $-0.01, -$ & $-$ & $-, +0.2$ & $-, +0.1$ & $-0.1, +0.1$ & $-0.01, -$ & $-0.01, +0.01$ \\
\midrule
\tch modelling & $-0.01, +0.01$ & $-$ & $-0.3, +0.1$ & $-$ & $-$ & $-0.04, +0.04$ & $-0.15, +0.15$ \\
\ttbar modelling & $-0.03, +0.02$ & $-0.01, -$ & $-0.5, +0.4$ & $-0.1, +0.1$ & $-0.1, +0.1$ & $-0.01, +0.03$ & $-0.04, +0.08$ \\
\Wjets modelling & $-0.01, +0.02$ & $-0.01, +0.01$ & $-0.8, +0.6$ & $-0.1, +0.1$ & $-, +0.1$ & $-0.01, +0.02$ & $-0.06, +0.08$ \\
\midrule
MC sample sizes & $-0.03, +0.03$ & $-0.03, +0.03$ & $-0.9, +0.8$ & $-0.2, +0.2$ & $-0.2, +0.2$ & $-0.03, +0.05$ & $-0.14, +0.14$ \\
Data sample size & $-0.06, +0.05$ & $-0.04, +0.04$ & $-3.1, +1.7$ & $-0.5, +0.7$ & $-0.5, +0.5$ & $-0.05, +0.08$ & $-0.13, +0.15$ \\
\midrule
Total systematic & $-0.07, +0.07$ & $-0.02, +0.02$ & $-1.8, +1.5$ & $-0.4, +0.5$ & $-0.5, +0.5$ & $-0.07, +0.10$ & $-0.35, +0.36$ \\
\midrule
\textbf{Total} & $-0.10, +0.09$ & $-0.06, +0.06$ & $-3.9, +2.6$ & $-0.8, +0.9$ & $-0.7, +0.7$ & $-0.09, +0.14$ & $-0.41, +0.43$ \\
\bottomrule
\end{tabular}%
}

\end{table}
\FloatBarrier

% End of text imported from the .//tables/ranking.tex input file

The effect of each systematic uncertainty on the WCs is evaluated by removing that systematic uncertainty from Eq.~(\ref{eq:systcov}) and comparing the $-\ln\mathcal{L}$ scan with that for the nominal fit.
Similarly, the impact of the MC statistical uncertainty is obtained by comparing the $-\ln\mathcal{L}$ scans with and without inclusion of $\mathbb{C}_\text{MC}^\text{stat.}$.
~\cref{table:uncertainty_summary} summarises the sources of both the systematic and statistical uncertainties for the $1\sigma$ confidence interval for each of the POIs.
Although the total systematic uncertainty is comparable
to the statistical uncertainty of the data except for
\cpq, no single source of systematic uncertainty significantly outweighs the
overall MC statistical uncertainty, which is distributed among the dominant simulated processes, with the $t$-channel contribution slightly larger than those from the $\mathrm{t\bar{t}}$ and $W$+jets samples, which are of comparable size.

\begin{figure}[!htb]
\begin{center}{\includegraphics[width=0.7\textwidth]{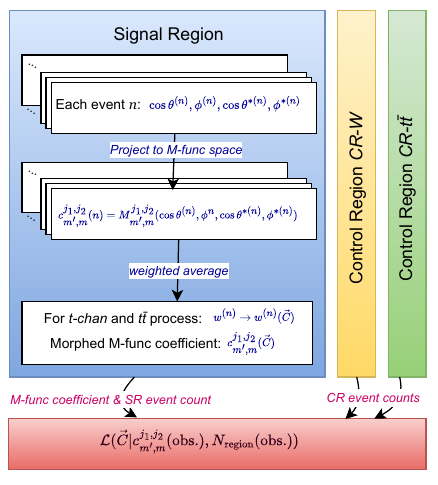}} \\
\caption{Schematic overview of the analysis workflow, from individual events selected in SR and CRs to the final likelihood function used for limit settings.}
\label{fig:fourangle_sketch}
\end{center}
\end{figure}
\FloatBarrier

Finally, the overall analysis workflow is illustrated in~\cref{fig:fourangle_sketch}.
For the SR, the four decay angles are reconstructed event by event
and inserted into Eq.~(\ref{ec:M}) to compute the coefficients
of the orthonormal \mfunc basis. These per-event coefficients are then combined via a
(weighted) average over all SR events (as defined in Eq.~(\ref{ec:M})) to form the final \mfunc observables.
The dependence of the (averaged) coefficients on the WC vector, $\vec{C}$, in both the \tch single-top-quark and \ttbar processes is obtained from MC simulation
and modelled using the morphing technique described in~\cref{sec:morphing}.
This approach fully captures the impact of the WCs on the quadruple-differential phase space without the need to bin a four-dimensional distribution.
A Gaussian likelihood $\mathcal{L}$, built from the averaged \mfunc coefficients and the event counts in the SR and CRs, is then maximised to extract the best-fit WC values.

% End of text imported from the .//sections/analysis.tex input file

%
%
%
%
\section{Results}
\label{sec:result}

% The next lines are included from the .//sections/results.tex input file
%

%

\begin{figure}[!htbp]
\begin{center}
\includegraphics[width=\textwidth]{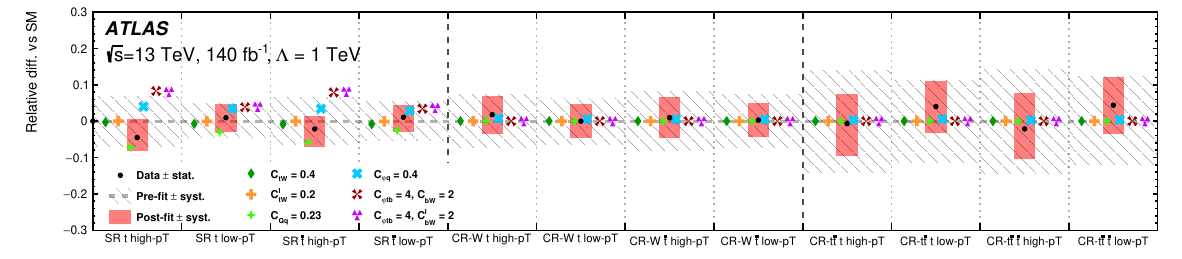}
\caption{The relative differences between the observed and SM predicted event counts in the SRs and CRs, for both the observed data and the fitted EFT model. The black points represent observed data, with the data statistical uncertainties as uncertainty bars. The grey dashed area indicates the total systematic uncertainties, while the red boxes are centred at positions determined by the best-fit values of the WCs, with heights determined by systematic uncertainties constrained by the fit. The coloured scattered markers correspond to expected values when making various WC assumptions.}
\label{fig:postfit_obs_evtcount}
\end{center}
\end{figure}

\begin{figure}[!htbp]
\begin{center}
\begin{subfigure}[b]{1.0\textwidth}
\includegraphics[width=\textwidth]{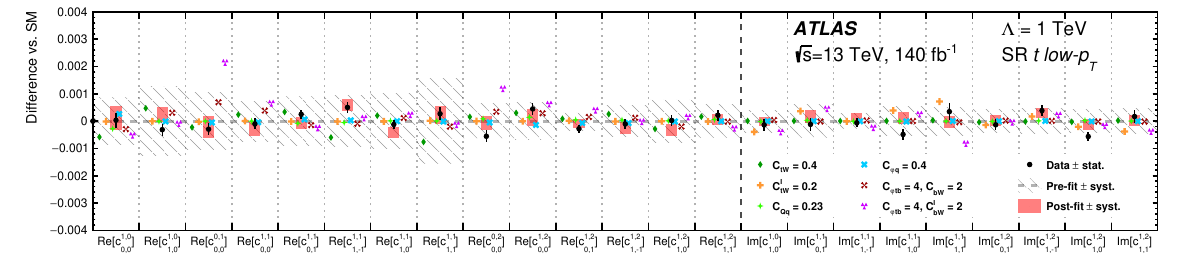}
\caption{}
\end{subfigure}
\begin{subfigure}[b]{1.0\textwidth}
\includegraphics[width=\textwidth]{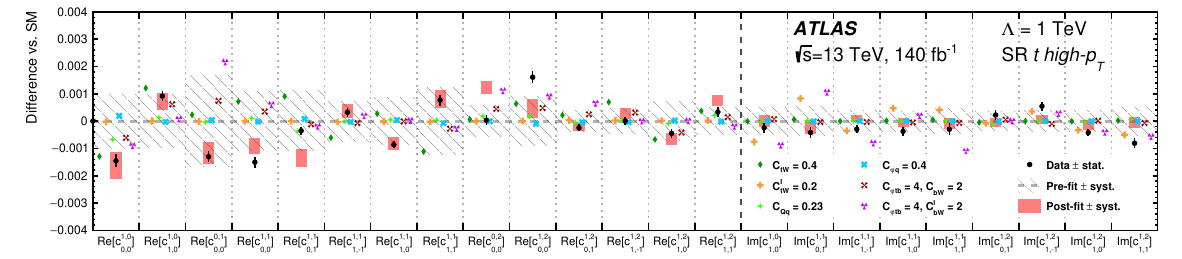}
\caption{}
\end{subfigure}
\begin{subfigure}[b]{1.0\textwidth}
\includegraphics[width=\textwidth]{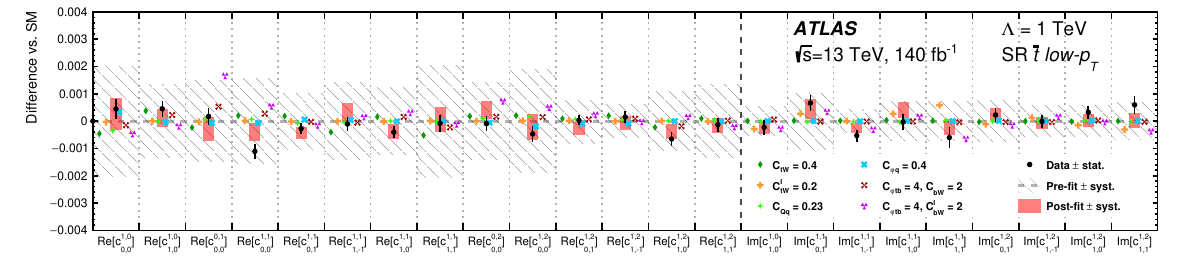}
\caption{}
\end{subfigure}
\begin{subfigure}[b]{1.0\textwidth}
\includegraphics[width=\textwidth]{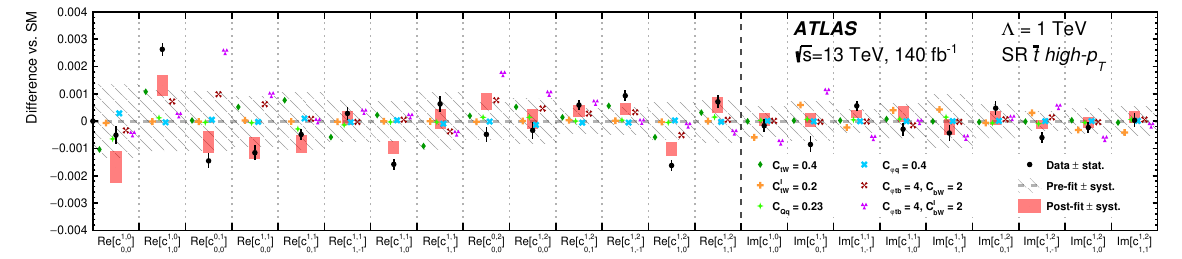}
\caption{}
\end{subfigure}
\caption{The absolute deviations of the \mfunc coefficients, as defined in Eq. (\ref{eq:mfcomp}), from the SM prediction, for both the observed data and the fitted EFT model, in the SR of the (a) low-\pT ($< 80$\,\GeV) top-quark channel, (b) high-\pT ($\geq 80$\,\GeV) top-quark channel, (c) low-\pT top-antiquark channel, and (d) high-\pT top-antiquark channel, respectively. The black points represent observed data, with the data statistical uncertainties as uncertainty bars. The grey dashed area indicates the total systematic uncertainties, while the red boxes are centred at positions determined by the best-fit values of the WCs, with heights determined by systematic uncertainties constrained by the fit. The coloured scattered markers correspond to expected values when making various WC assumptions.}
\label{fig:postfit_obs_mfunc}
\end{center}
\end{figure}

Summing over the four channels, the final observables are the $3\times4$ event counts in the SR and two CRs, along with the $23\times4$ \mfunc coefficients in the SR.
\cref{fig:postfit_obs_evtcount} shows the relative differences between the observed event counts and the SM predictions.~\cref{fig:postfit_obs_mfunc} shows the absolute deviations of the \mfunc coefficients from the SM prediction, for both the observed data and the fitted EFT model in each channel's SR. The black points represent the data, while the scattered markers correspond to the expected values when making various WC assumptions. The grey band, centred at zero, represents the SM expectation; its width indicates the pre-fit systematic uncertainties. The centre of each red box is determined by the best-fit value from the EFT fit; its height is set by the post-fit uncertainties, which include the effect of correlations. The plots demonstrate the effect of the data upon the fitted model.

\FloatBarrier

As expected, due to the large number of observables compared to the number of unconstrained parameters, the uncertainties of the observables are reduced by the fit. %
The SM prediction is in reasonable overall agreement with the 104 observables, with a $\chi^2=-2 \ln \mathcal{L}=105.2$, indicating a p-value at 44.9\%, although agreement is poorer for some high-\pT SR observables.
At the best-fit point of the EFT fit with nine parameters, the likelihood has a $\chi^2=87.4$ for 95 degrees of freedom (p-value of 70.0\%).  Both hypotheses are consistent with the data.

\FloatBarrier

\begin{figure}[!h]
\begin{center}
\begin{subfigure}[b]{0.38\textwidth}
\includegraphics[width=\textwidth]{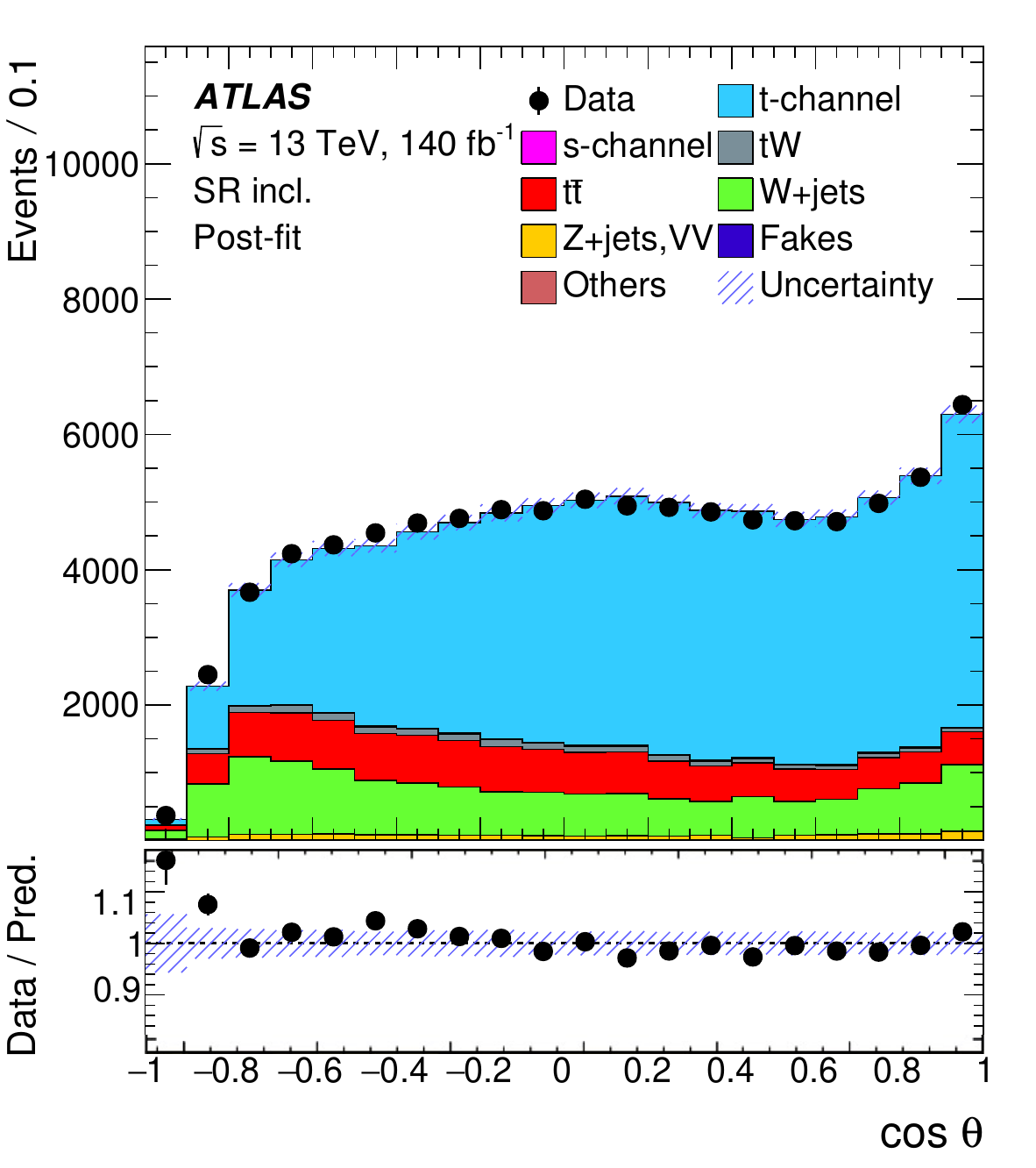}
\caption{}
\end{subfigure}
\hspace{0.08\textwidth}
\begin{subfigure}[b]{0.38\textwidth}
\includegraphics[width=\textwidth]{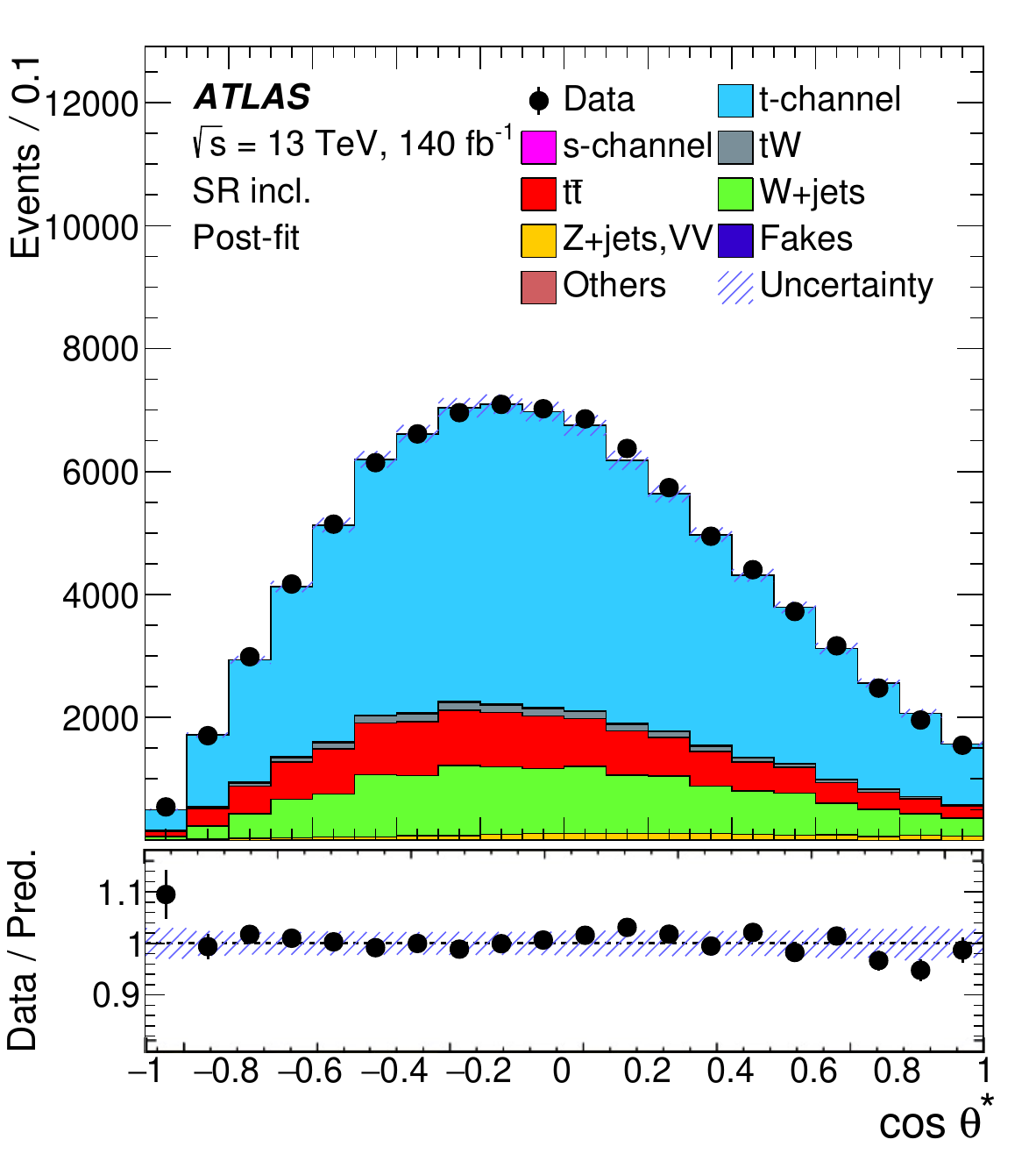}
\caption{}
\end{subfigure}
\begin{subfigure}[b]{0.38\textwidth}
\includegraphics[width=\textwidth]{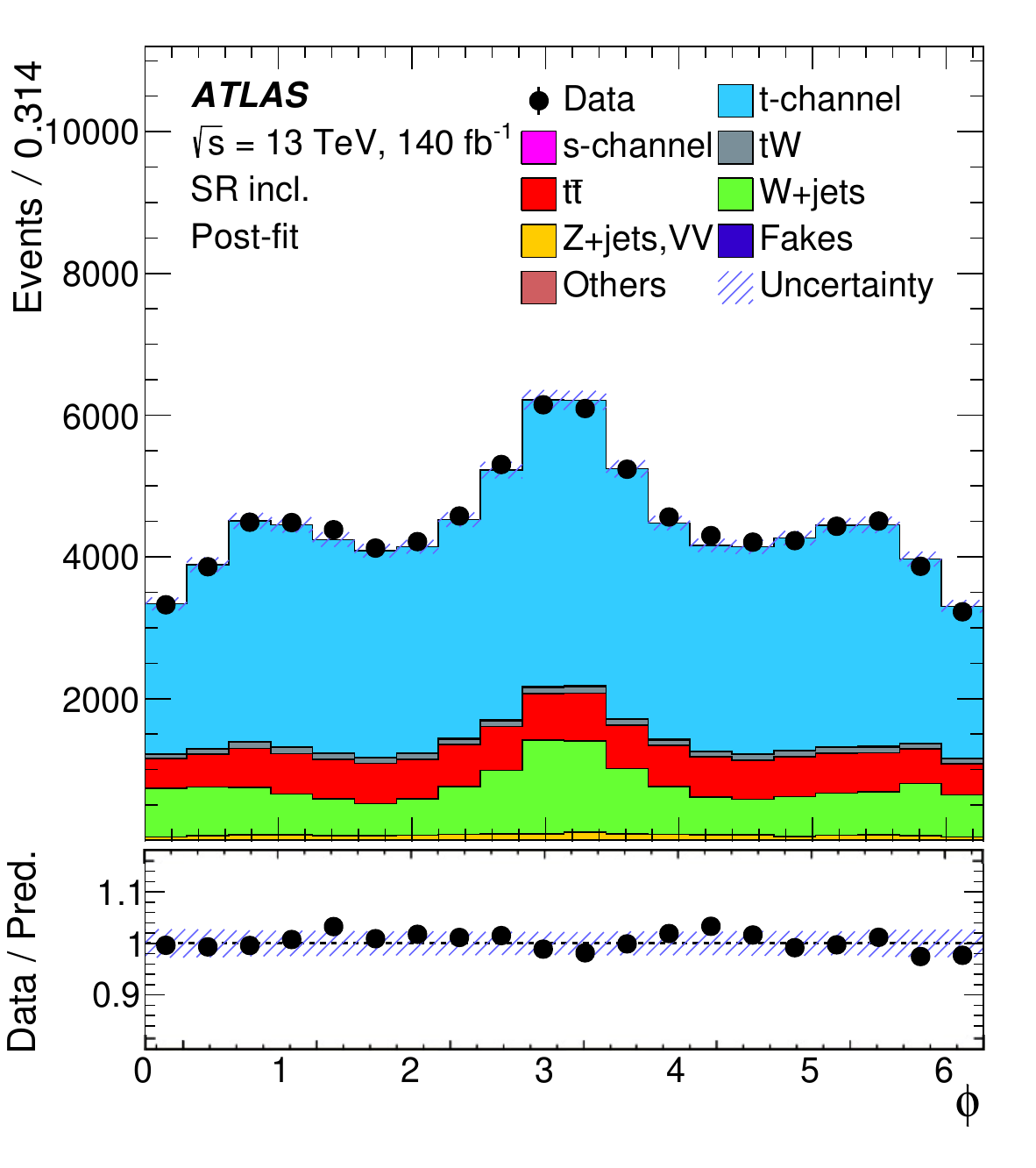}
\caption{}
\end{subfigure}
\hspace{0.08\textwidth}
\begin{subfigure}[b]{0.38\textwidth}
\includegraphics[width=\textwidth]{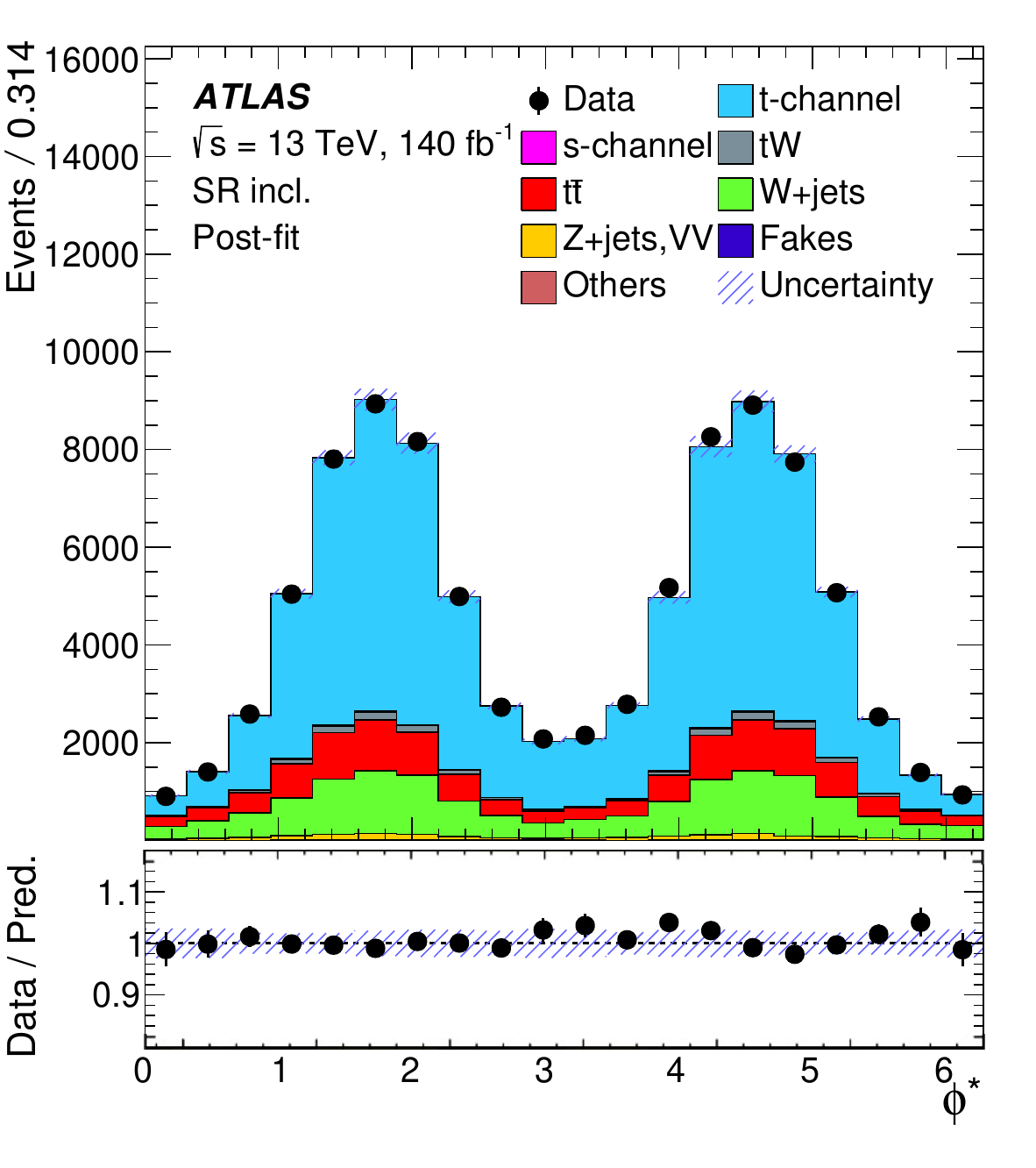}
\caption{}
\end{subfigure}
\caption{Post-fit distributions of the four angles, namely (a) $\cos\theta$, (b) $\cos\theta^{*}$, (c) $\phi$, and (d) $\phi^{*}$ in the inclusive SR. The fitted predicted distributions are compared with the observed data, depicted as black points with statistical uncertainties. The uncertainty bands reflect the statistical and systematic uncertainties, taking into account the post-fit constraints from the profile likelihood fit. Correlations among all parameters obtained from the fit are accounted for in the calculation of the uncertainty bands. The lower plot for each angular distribution shows the ratio of the observed data to the fitted prediction in each bin, offering a quantitative measure of how well the two agree.}
\label{fig:postfit_angles}
\end{center}
\end{figure}

The best-fit values of the WCs are substituted into the morphing functions of the \tch and \ttbar cross-sections to obtain the post-fit angular distributions, in order to demonstrate that the overall agreement between data and MC is preserved after the fit.
The post-fit SR distributions of the four angles $\cos\theta$, $\cos\theta^{*}$, $\phi$, and $\phi^{*}$ are shown in \cref{fig:postfit_angles}. They are all in good agreement with the data, with $\chi^2$ values ranging from 5 to 18 for 20 bins.

Three different fitting schemes are considered for a one-dimensional likelihood scan:
\begin{itemize}
\item \textbf{Fully profiled fit}: each POI is scanned while profiling the other six POIs.
\item \textbf{Reduced fit}: WCs that are dominated by quadratic dependencies (\cptb, \cbw, \cibw) are fixed at SM values in this scheme. This fit allows one to obtain well-defined point estimates and a covariance matrix for the remaining parameters.
\item \textbf{Individual fit}: all other POIs fixed at the SM values when scanning a POI.
\end{itemize}
The fitted scale factors of the two major background processes \ttbar and \Wjets are summarised in \cref{table:unblind_sf}.

% The next lines are included from the .//tables/unblind_sf.tex input file
\begin{table}[htbp]
\centering
\caption{The best-fit results for the scale-factor vector $\vec{S}$ with the corresponding $68\%$ CL intervals.}
\resizebox{0.84\textwidth}{!}{%
\begin{tabular}{@{}l c S[table-format=7.5] c S[table-format=5.5] c S[table-format=7.5] c S[table-format=5.5] @{}}
\toprule
\multirow{2}{*}{\textbf{Scale factor}} & \multicolumn{2}{c}{Fully profiled fit} & \multicolumn{2}{c}{Reduced fit} \\
& \textbf{Best-fit value} & \textbf{\(68\%\) CL interval} & \textbf{Best-fit value} & \textbf{\(68\%\) CL interval} \\
\midrule
\(S^\text{\Wjets}\) & 0.92 & \pm0.07  & 0.91 & \pm0.06\\ %
\(S^\text{\ttbar}\) & 1.09 & \pm0.08  & 1.10 & \pm0.08 \\ %
\bottomrule
\end{tabular}
}
\label{table:unblind_sf}
\end{table}
\FloatBarrier

% End of text imported from the .//tables/unblind_sf.tex input file

%
%
%

%
%

%
%
%
%
%
%
%
%
%
%
%
%
%
%
%
%
%
%
%
%
%
%
%
%
%
%
%
%
%
%
%
%
%
%
%
%
%
%
%
%
%
%
%
%
%
%
%
%
%
%
%
%
%

%
%
%

\begin{figure}[!htbp]
\begin{center}
\begin{subfigure}[b]{0.41\textwidth}
\includegraphics[width=\textwidth]{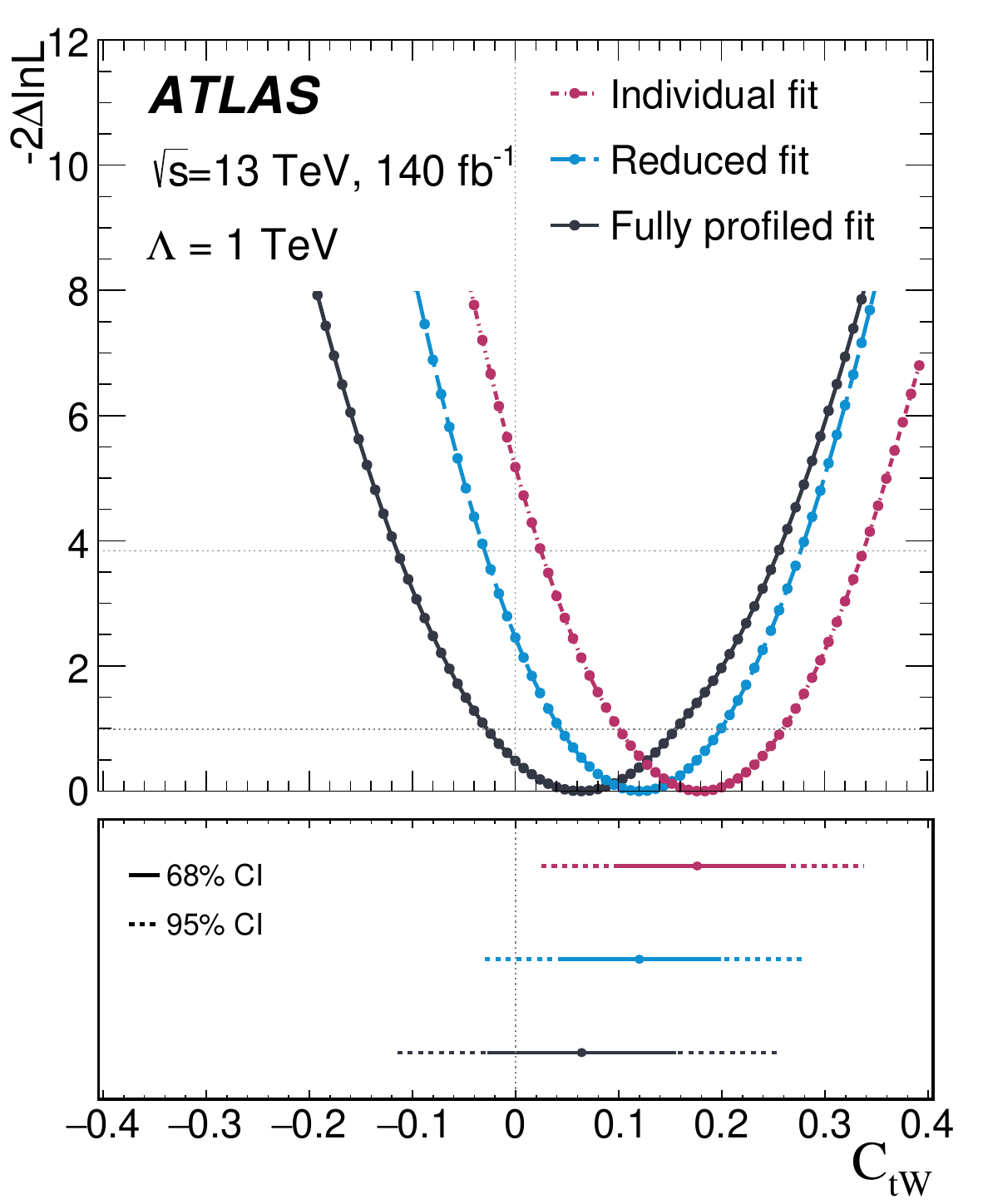}
\caption{}
\end{subfigure}
\hspace{0.08\textwidth}
\begin{subfigure}[b]{0.41\textwidth}
\includegraphics[width=\textwidth]{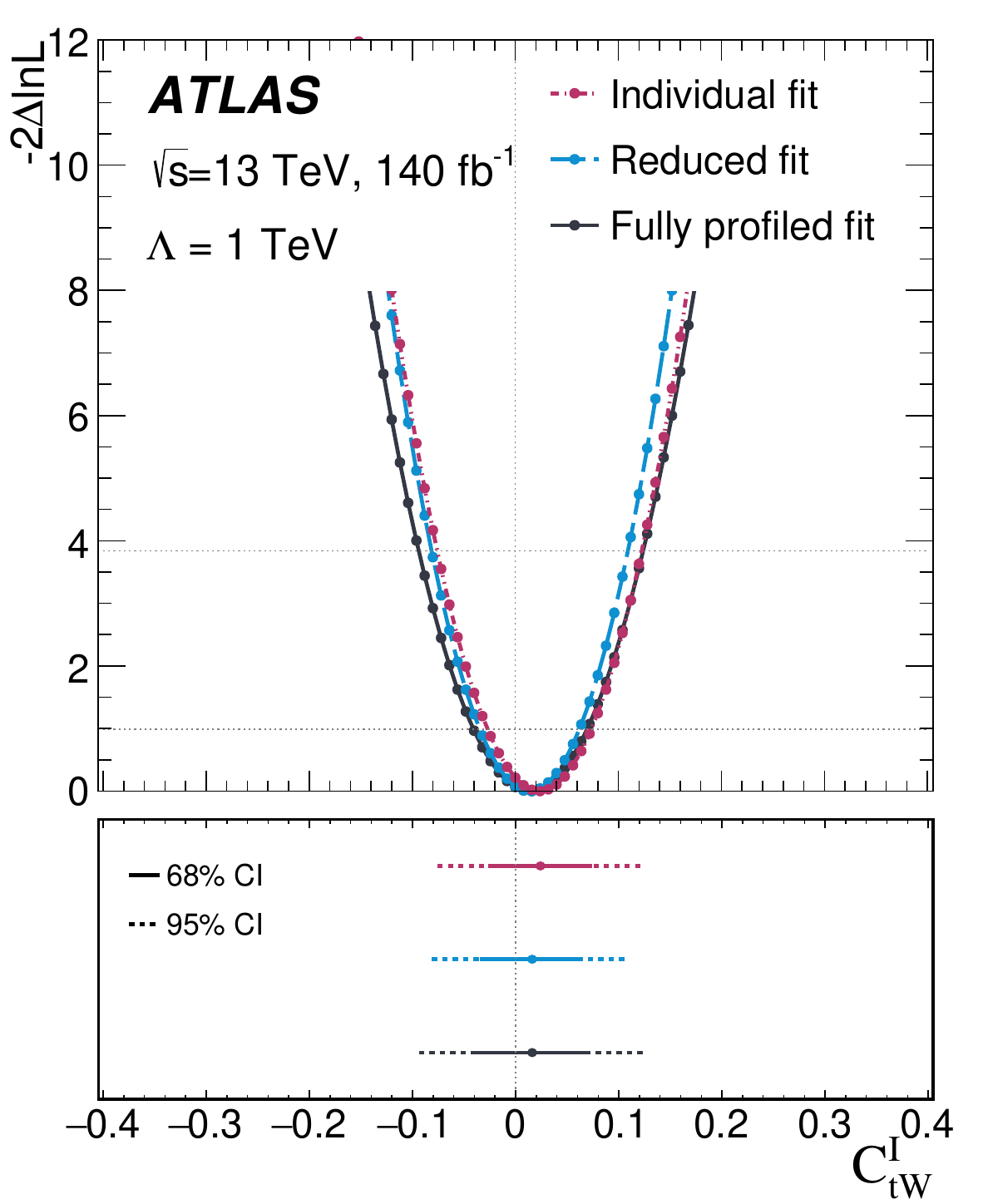}
\caption{}
\end{subfigure}
\begin{subfigure}[b]{0.41\textwidth}
\includegraphics[width=\textwidth]{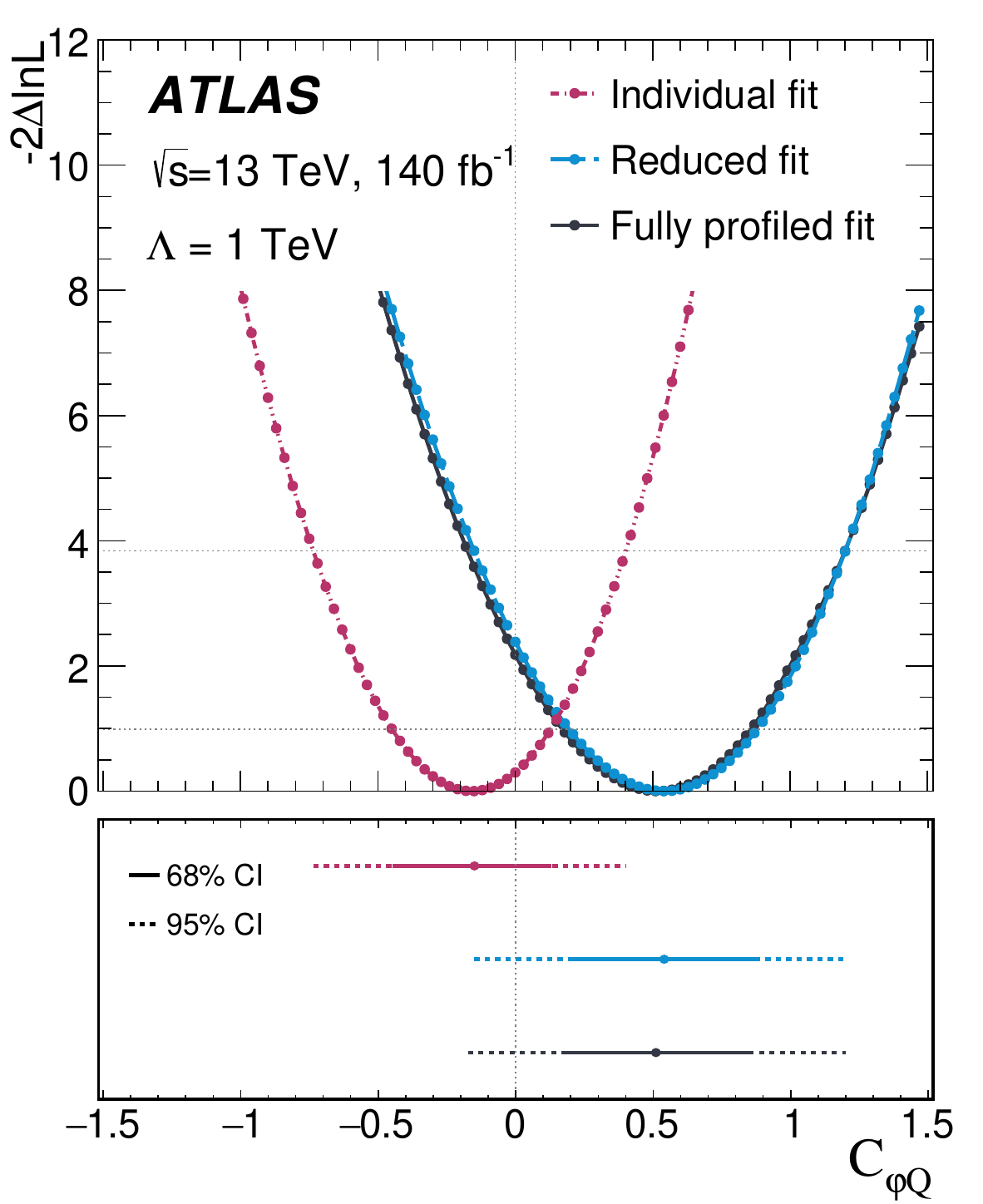}
\caption{}
\end{subfigure}
\hspace{0.08\textwidth}
\begin{subfigure}[b]{0.41\textwidth}{\includegraphics[width=\textwidth]{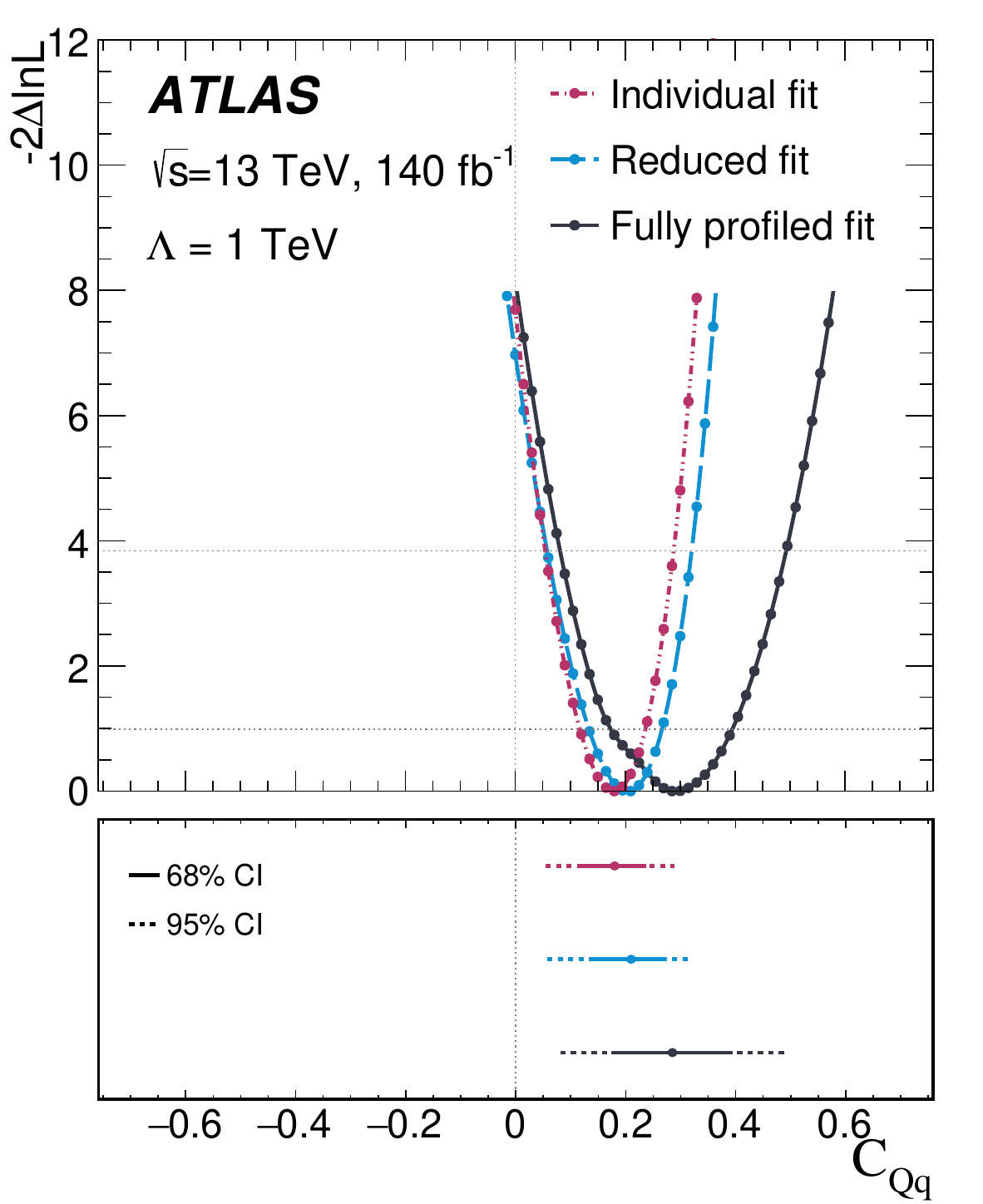}}
\caption{}
\end{subfigure}

\caption{The likelihood scans for the parameters (a) \ctw, (b) \citw, (c) \cpq, and (d) $\cqq$, comparing the fully profiled fit scheme (black curve), the reduced fit scheme (dashed blue curve) and the individual fit scheme (dash-dotted red curve). The 68\% and 95\% CL intervals (CI) for each parameter in each scheme are shown in the bottom plots.}
\label{fig:result_unblind_linear_only}
\end{center}
\end{figure}

\begin{figure}[!htbp]
\begin{center}
\begin{subfigure}[b]{0.46\textwidth}
\includegraphics[width=\textwidth]{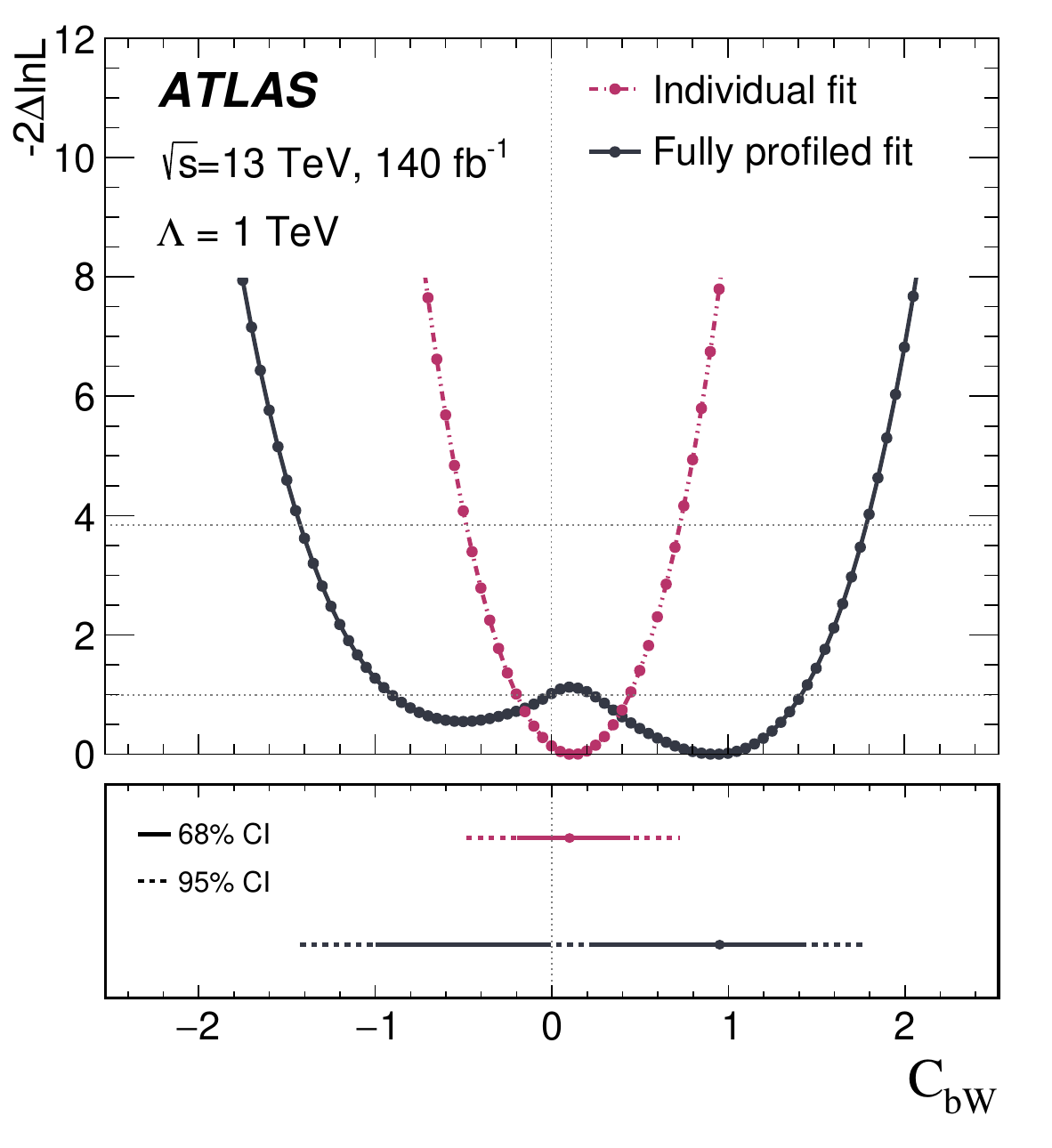}
\caption{}
\end{subfigure}
\hspace{0.08\textwidth}
\begin{subfigure}[b]{0.46\textwidth}
\includegraphics[width=\textwidth]{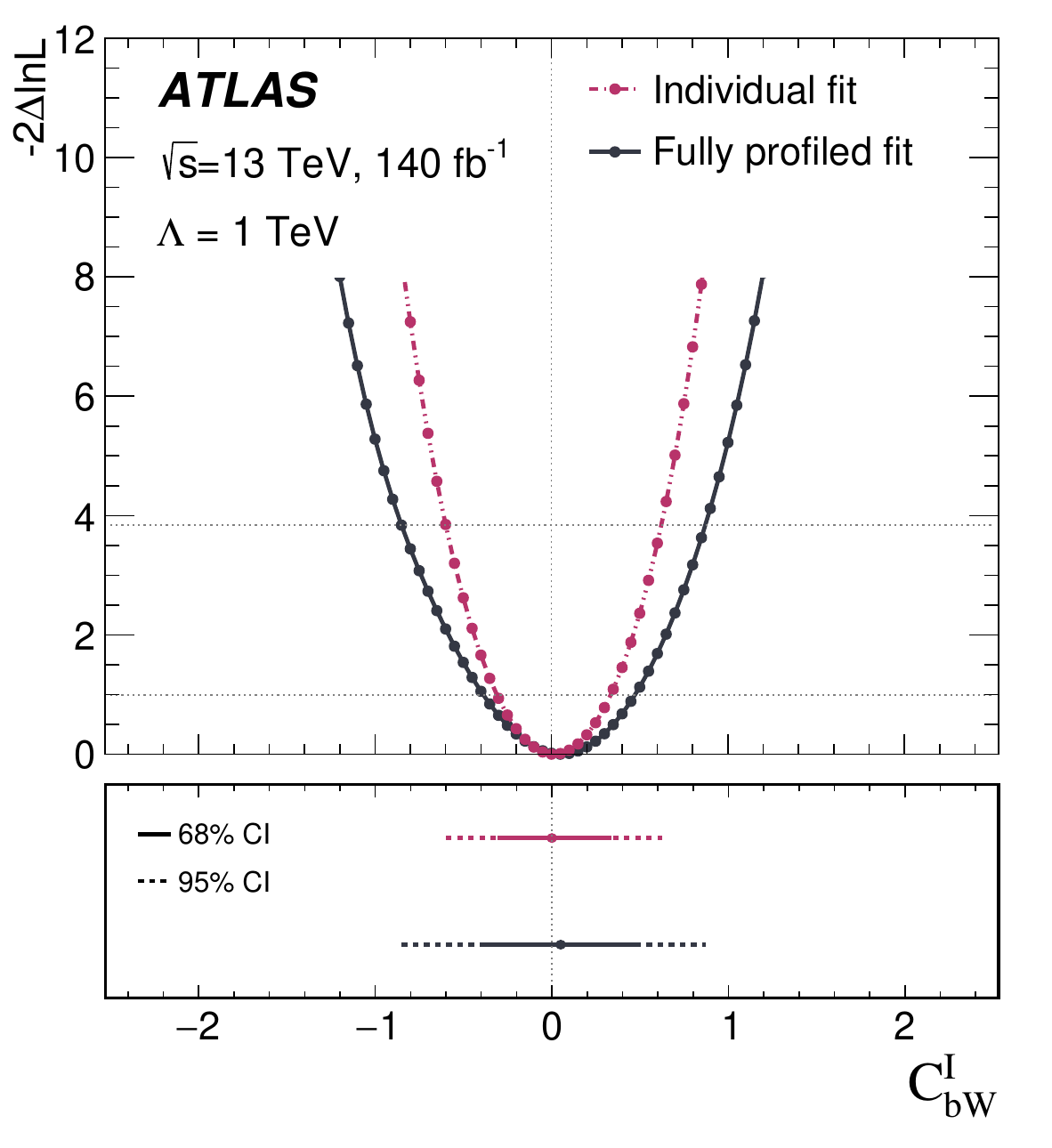}
\caption{}
\end{subfigure}
\begin{subfigure}[b]{0.46\textwidth}
\includegraphics[width=\textwidth]{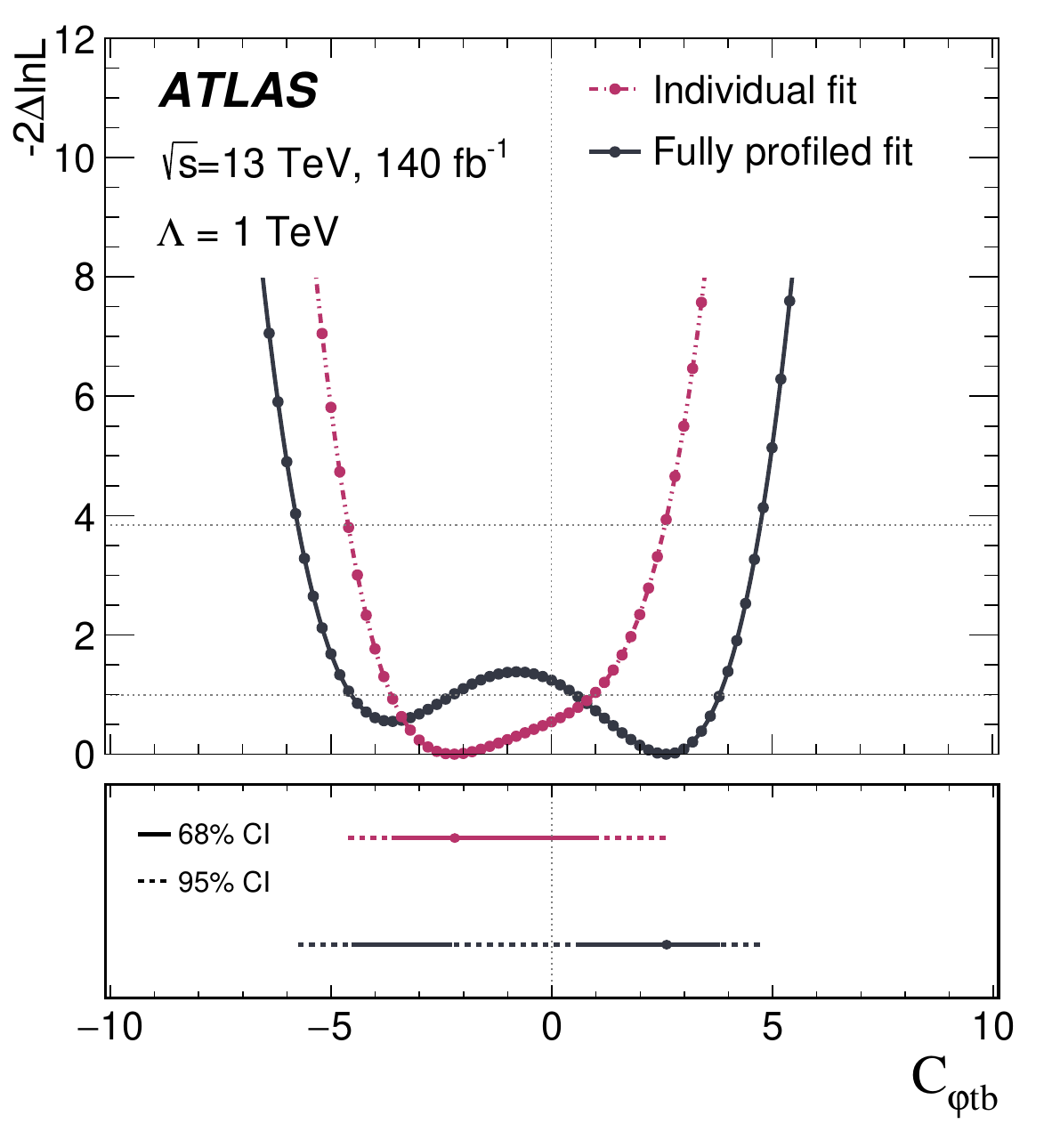}
\caption{}
\end{subfigure}

\caption{The likelihood scans for the parameters (a) \cbw, (b) \cibw and (c) \cptb, comparing the fully profiled fit scheme (black curve) with the individual fit scheme (dash-dotted red curve).
The 68\% and 95\% CL intervals (CI) for each parameter in each scheme are shown in the bottom plots.}
\label{fig:result_unblind_quadratic_only}
\end{center}
\end{figure}

One-dimensional likelihood scans for these fitting schemes are shown in \cref{fig:result_unblind_linear_only} for the EFT coefficients that have a linear effect on the \mfunc coefficients, and in \cref{fig:result_unblind_quadratic_only} for those that have a quadratic effect.
The non-zero value of the $b$-quark mass used in the generation of the EFT grid points used for the morphing leads to a slight linear dependence, so the resulting likelihood scans are not expected to be completely symmetric about the SM point.

\begin{figure}[!htbp]
\begin{center}{\hspace{0.0\textwidth}\includegraphics[height=0.39\textheight]{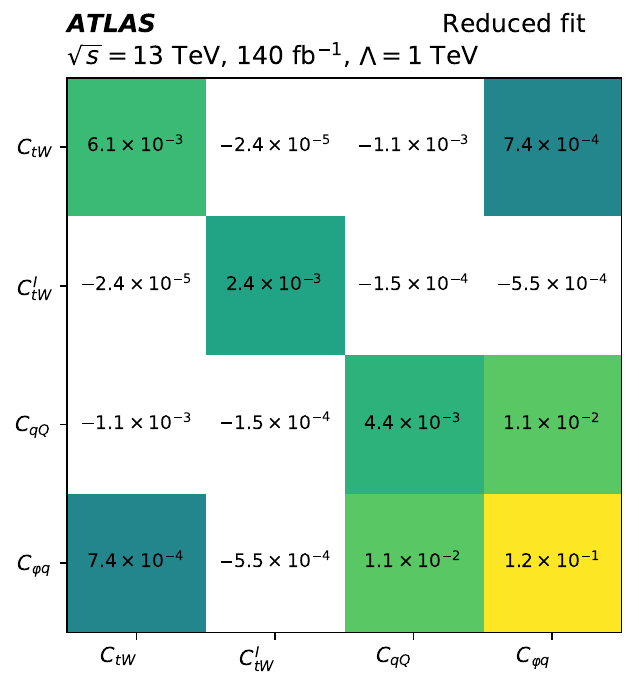}}
\caption{Covariance matrix for the reduced fit, assuming $\Lambda=1$ \TeV.}
\label{fig:reduced_fit_cov}
\end{center}
\end{figure}

The limits on each WC for each fitting scheme introduced above are summarised in~\cref{table:unblind_results_linear} and~\cref{table:unblind_results}. For the reduced fit, the covariance matrix
of the four WCs \ctw, \citw, \cqq and \cphiq is given in \cref{fig:reduced_fit_cov}.

\FloatBarrier

\FloatBarrier

% The next lines are included from the .//tables/limits_ub_linear.tex input file
\begin{table}[htbp]
\centering
\caption{The best-fit values obtained by the reduced fit scheme, and the 68\% and 95\% CL intervals. In this fitting scheme, only the WCs \ctw, \citw, \cqq and \cphiq that have a linear effect on the \mfunc coefficients are free floating. }
\resizebox{0.65\textwidth}{!}{%
\begin{tabular}{@{}c| c c c c c c c c@{}}
\toprule
\textbf{Wilson coefficient} & \multicolumn{3}{c}{\textbf{Reduced fit}}  \\
\([\TeV^{-2}]\) & \(\xspace\)Central value\(\xspace\) & \(68\%\) CL interval & \(95\%\) CL interval\\
\midrule
\(\ctw/\Lambda^2\) & \(0.12\) & \([+0.04, +0.20]\) & \([-0.03, +0.28]\) \\ \addlinespace
\(\citw/\Lambda^2\) & \(0.02\)  & \([-0.03, +0.06]\) & \([-0.08, +0.11]\) \\ \addlinespace
\(\cqq/\Lambda^2\) & \(0.21\)  & \([+0.13, +0.27]\) & \([+0.05, +0.32]\) \\ \addlinespace
\(\cphiq/\Lambda^2\) & \( 0.54 \)  & \([+0.19, +0.88]\) & \([-0.16, +1.21]\) \\ \addlinespace
\bottomrule
\end{tabular}
}
\label{table:unblind_results_linear}
\end{table}
\FloatBarrier

% End of text imported from the .//tables/limits_ub_linear.tex input file

% The next lines are included from the .//tables/limits_table_ub.tex input file
\begin{table}[htbp]
\centering
\caption{The WC intervals at $68\%$ and $95\%$ CL, obtained by a one-dimensional likelihood scan. The table compares two scenarios: a fully profiled fit in which all non-scanned WCs are profiled simultaneously, and an individual fit in which all other coefficients are kept fixed at their SM value (zero).}
\resizebox{0.95\textwidth}{!}{%
\begin{tabular}{@{}c | c c | c c  c c c c@{}}
\toprule
\textbf{Wilson coefficient} & \multicolumn{2}{c|}{\textbf{Fully profiled fit}}  & \multicolumn{2}{c}{\textbf{Individual fit}} \\
\([\TeV^{-2}]\) & \(68\%\) CL interval & \(95\%\) CL interval & \(68\%\) CL interval & \(95\%\) CL interval \\
\midrule
\multicolumn{1}{c|}{\(\ctw/\Lambda^2\)} & \([-0.03, +0.17]\) & \([-0.12, +0.26]\)  & \([+0.10, +0.26]\) & \([+0.02, +0.34]\) \\ \addlinespace
\multicolumn{1}{c|}{\(\citw/\Lambda^2\)} & \([-0.04, +0.07]\) & \([-0.10, +0.13]\) \ & \([-0.03, +0.07]\) & \([-0.08, +0.12]\) \\ \addlinespace
\multicolumn{1}{c|}{\(\cptb/\Lambda^2\)} & \([-4.6, -2.3] \cup [+0.5, +3.8]\) & \([-5.8, +4.8]\) & \([-3.6, +1.0]\) & \([-4.6, +2.6]\) \\ \addlinespace
\multicolumn{1}{c|}{\(\cbw/\Lambda^2\)} & \([-1.0, 0.0] \cup [+0.2, +1.4]\) & \([-1.4, +1.8]\) & \([-0.2, +0.4]\) & \([-0.5, +0.7]\) \\ \addlinespace
\multicolumn{1}{c|}{\(\cibw/\Lambda^2\)} & \([-0.4, +0.5]\) & \([-0.9, +0.9]\)  & \([-0.3, +0.3]\) & \([-0.6, +0.6]\) \\ \addlinespace
\multicolumn{1}{c|}{\(\cqq/\Lambda^2\)} & \([+0.17, +0.37]\) & \([+0.08, +0.50]\)  & \([+0.12, +0.24]\) & \([+0.05, +0.29]\) \\ \addlinespace
\multicolumn{1}{c|}{\(\cphiq/\Lambda^2\)} & \([+0.17, +0.86]\) & \([-0.19, +1.2]~~\)  & \([-0.45, +0.13]\) & \([-0.75, +0.41]\) \\ \addlinespace

\bottomrule
\end{tabular}
}
\label{table:unblind_results}
\end{table}
\FloatBarrier

% End of text imported from the .//tables/limits_table_ub.tex input file

\FloatBarrier

\begin{figure}[!htbp]
\begin{center}
\begin{subfigure}[b]{0.48\textwidth}
\includegraphics[width=\textwidth]{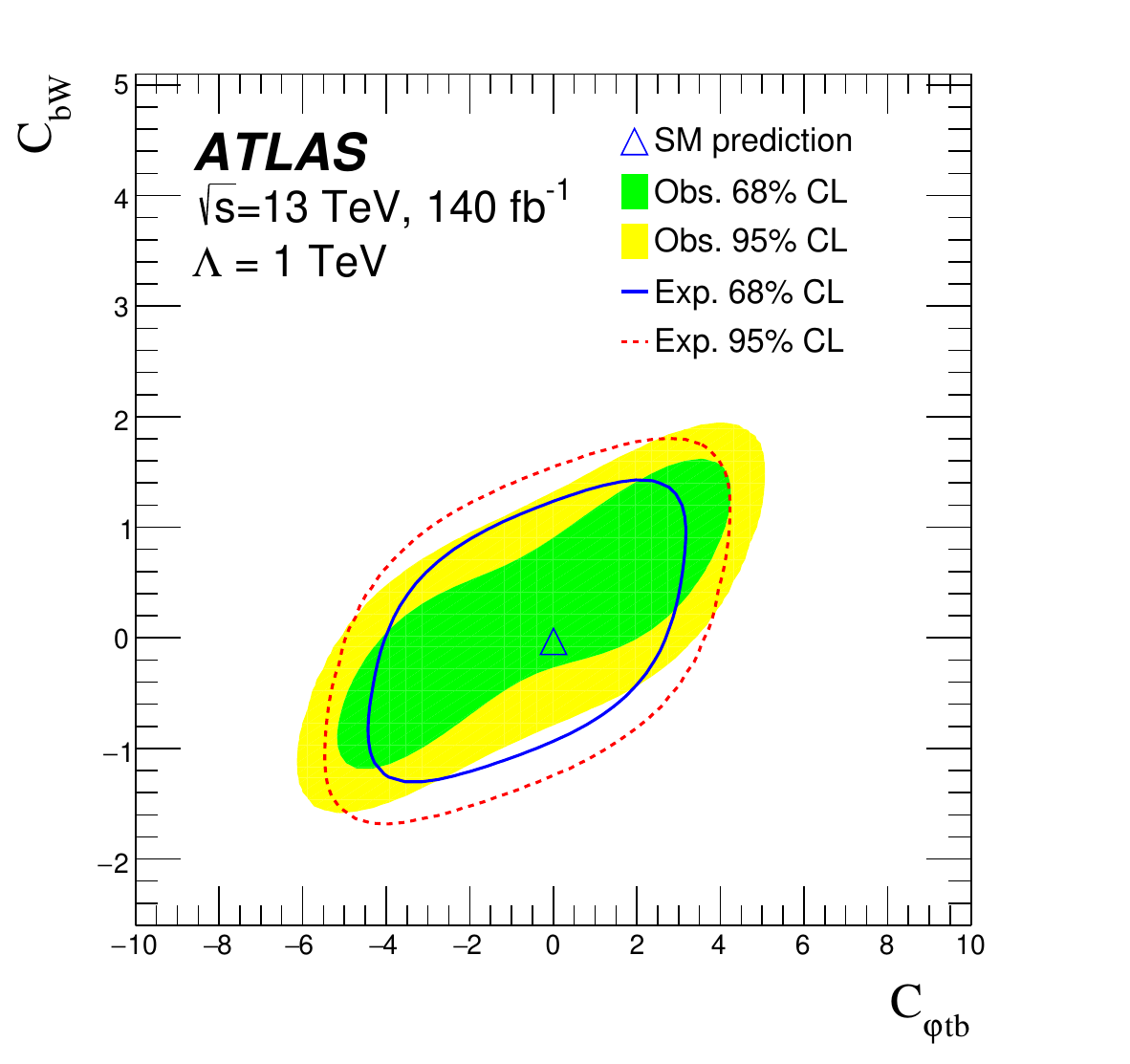}
\caption{}
\end{subfigure}
\xspace
\begin{subfigure}[b]{0.48\textwidth}
\includegraphics[width=\textwidth]{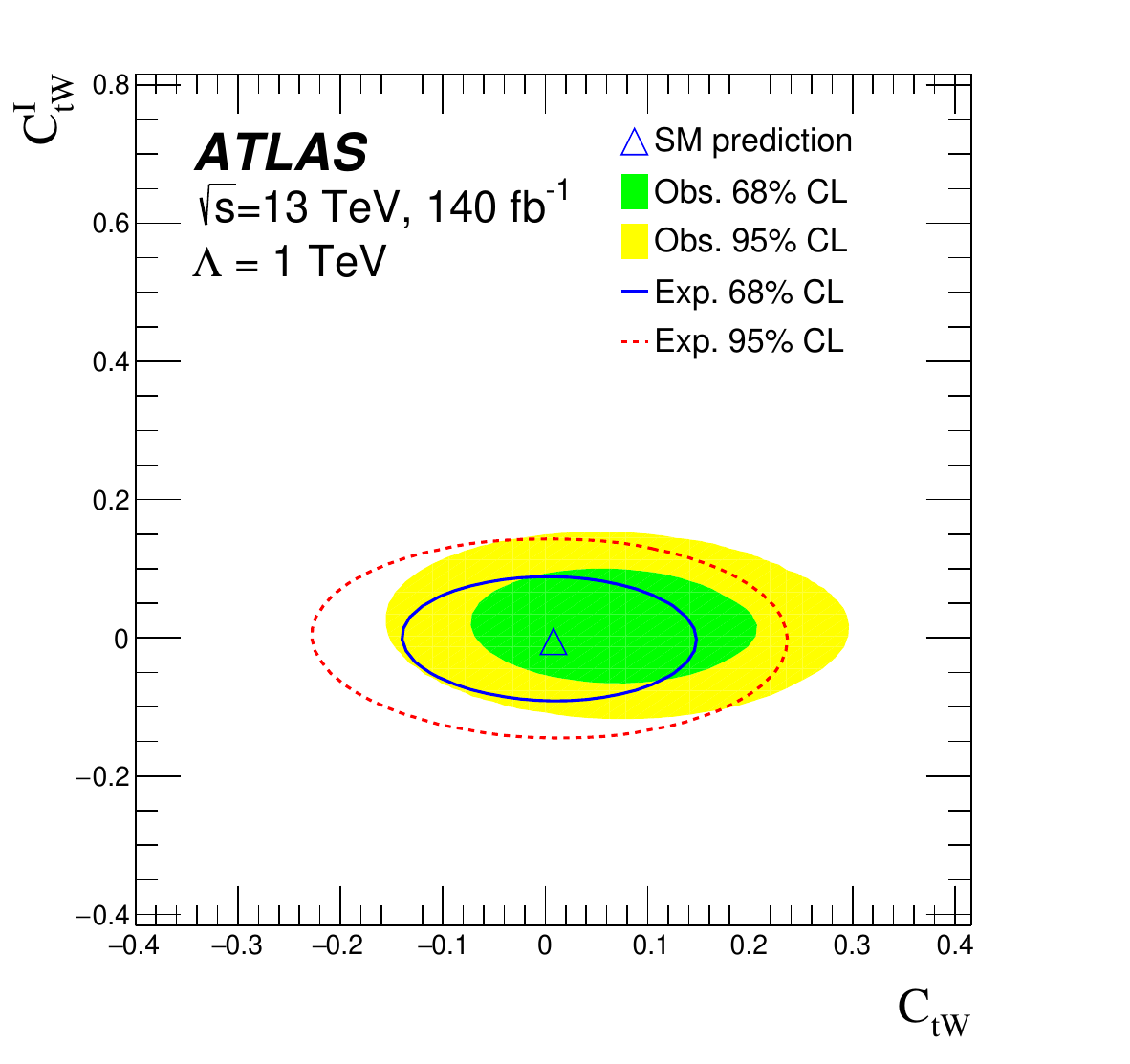}
\caption{}
\end{subfigure}
\caption{Likelihood scan of (a) \cptb and \cbw, and (b) \ctw and \citw, with all other parameters profiled. The expected SM point is marked with a blue triangle. The observed (expected) $68\%$ and $95\%$ CL regions are shown as inner green (blue) and outer yellow (red) regions (lines), respectively. }
\label{fig:result_unblind_cptb_cbw}
\end{center}
\end{figure}
\FloatBarrier
\begin{figure}[!htbp]
\begin{center}
\begin{subfigure}[b]{0.48\textwidth}
\includegraphics[width=\textwidth]{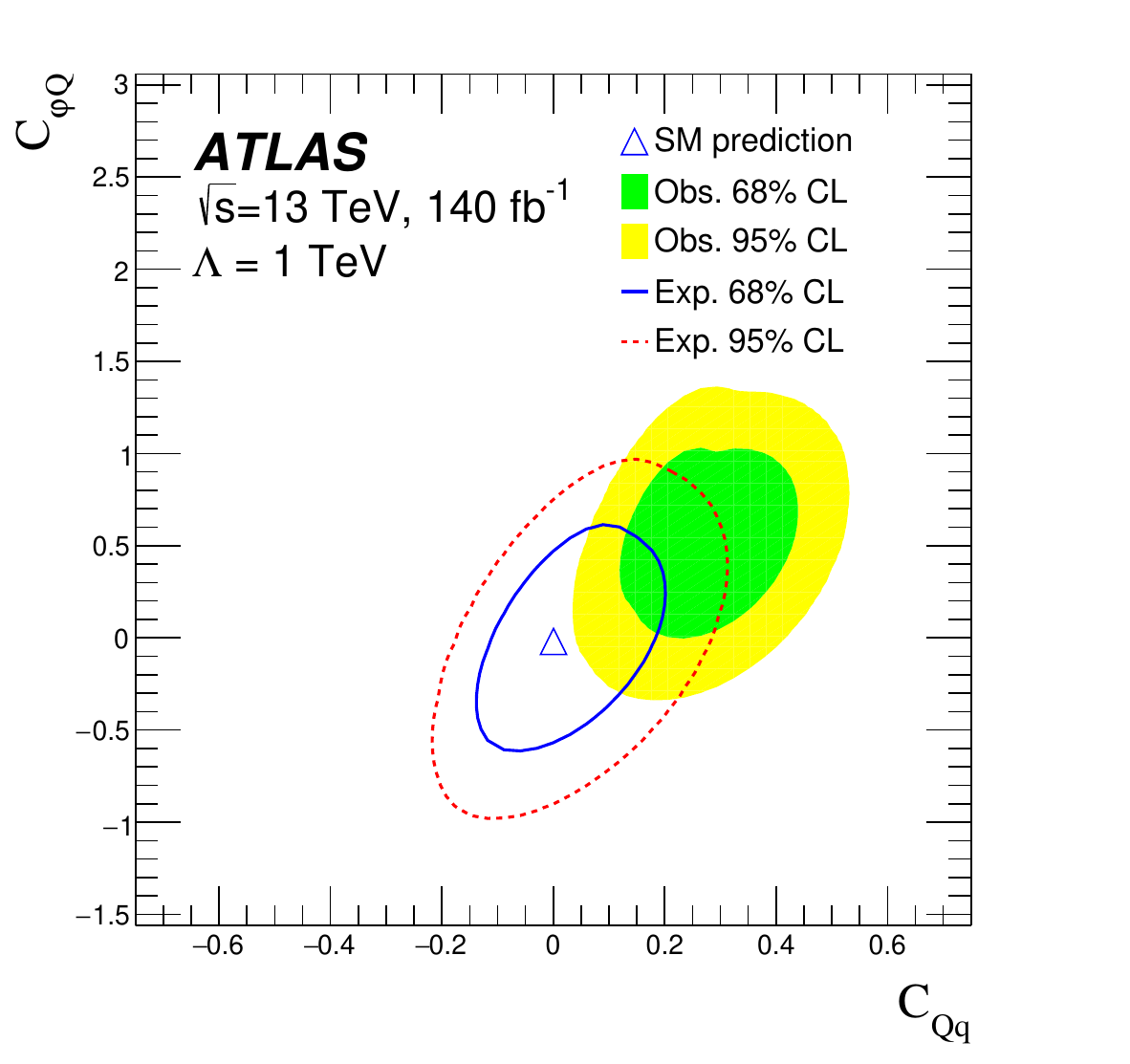}
\caption{}
\end{subfigure}
\begin{subfigure}[b]{0.48\textwidth}
\includegraphics[width=\textwidth]{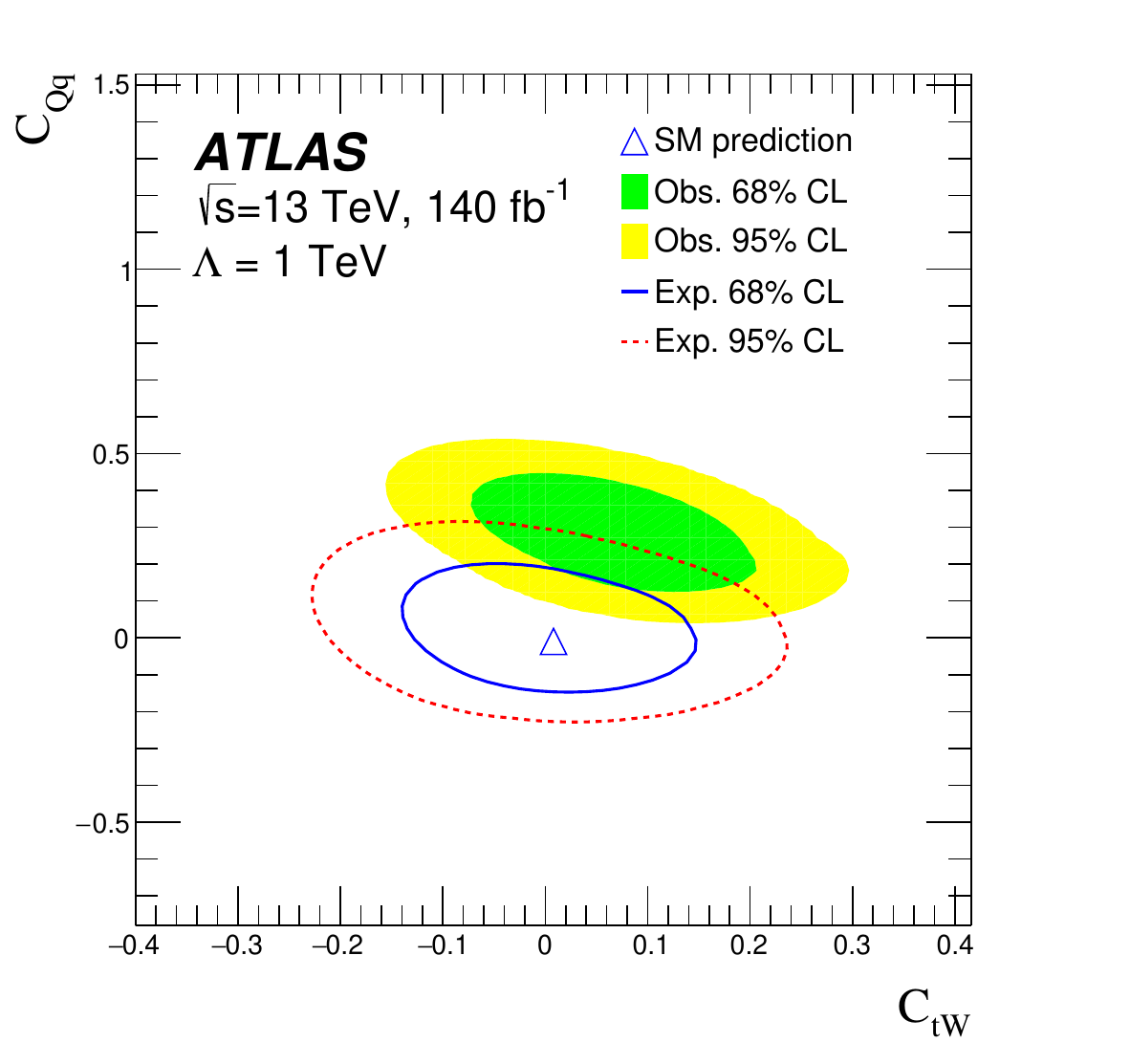}
\caption{}
\end{subfigure}
\caption{Likelihood scan of (a) \cqq and $\cpq$, and (b) \ctw and \cqq, with all other parameters profiled. The expected SM point is marked with a blue triangle. The observed (expected) $68\%$ and $95\%$ CL regions are as inner green (blue) and outer yellow (red) regions (lines), respectively. }
\label{fig:result_unblind_cqq_SFtchan}
\end{center}
\end{figure}
\FloatBarrier

Two-dimensional likelihood scans for the fully profiled fitting scheme are shown in \cref{fig:result_unblind_cptb_cbw} and \cref{fig:result_unblind_cqq_SFtchan}. The $68\%$ and $95\%$ CL regions are obtained by using the conventional $-2\Delta\ln \mathcal{L}$ thresholds for two degrees of freedom~\cite{ParticleDataGroup:2024cfk}.
It is known that one can construct a linear combination of
$\mathcal{O}_{\varphi tb}$ and $\mathcal{O}_{bW}$ that has no
sensitivity to the fraction of top quarks that decay to right-handed \PW bosons~\cite{Aguilar-Saavedra:2017wpl}. Additional observables in this analysis somewhat break that degeneracy, but a strong correlation still remains.
Negligible correlation between \ctw and \citw is observed.
There is, however, some correlation between the fitted results for \cqq and \cpq or \ctw.

\begin{figure}[!htb]
\begin{center}
\begin{subfigure}[b]{0.65\textwidth}
\includegraphics[width=\textwidth]{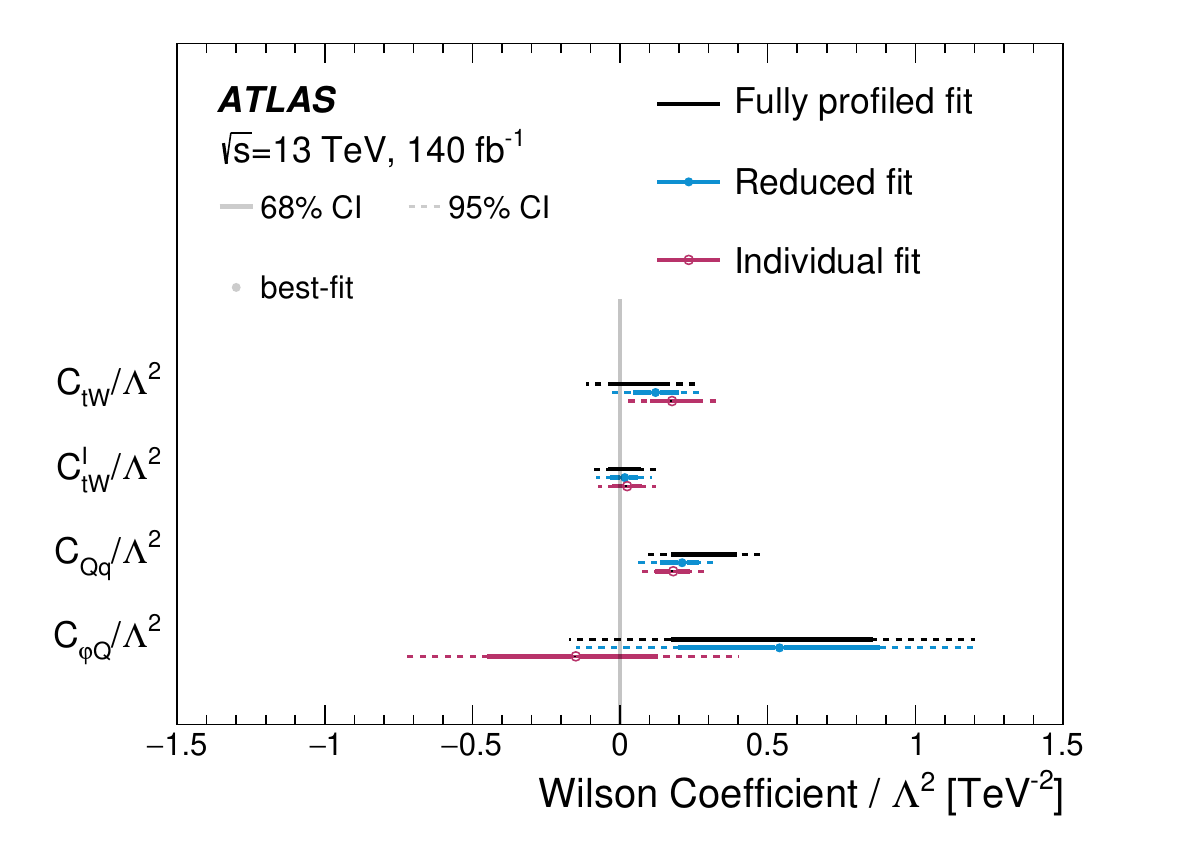}
\caption{}
\end{subfigure}
\begin{subfigure}[b]{0.65\textwidth}
\includegraphics[width=\textwidth]{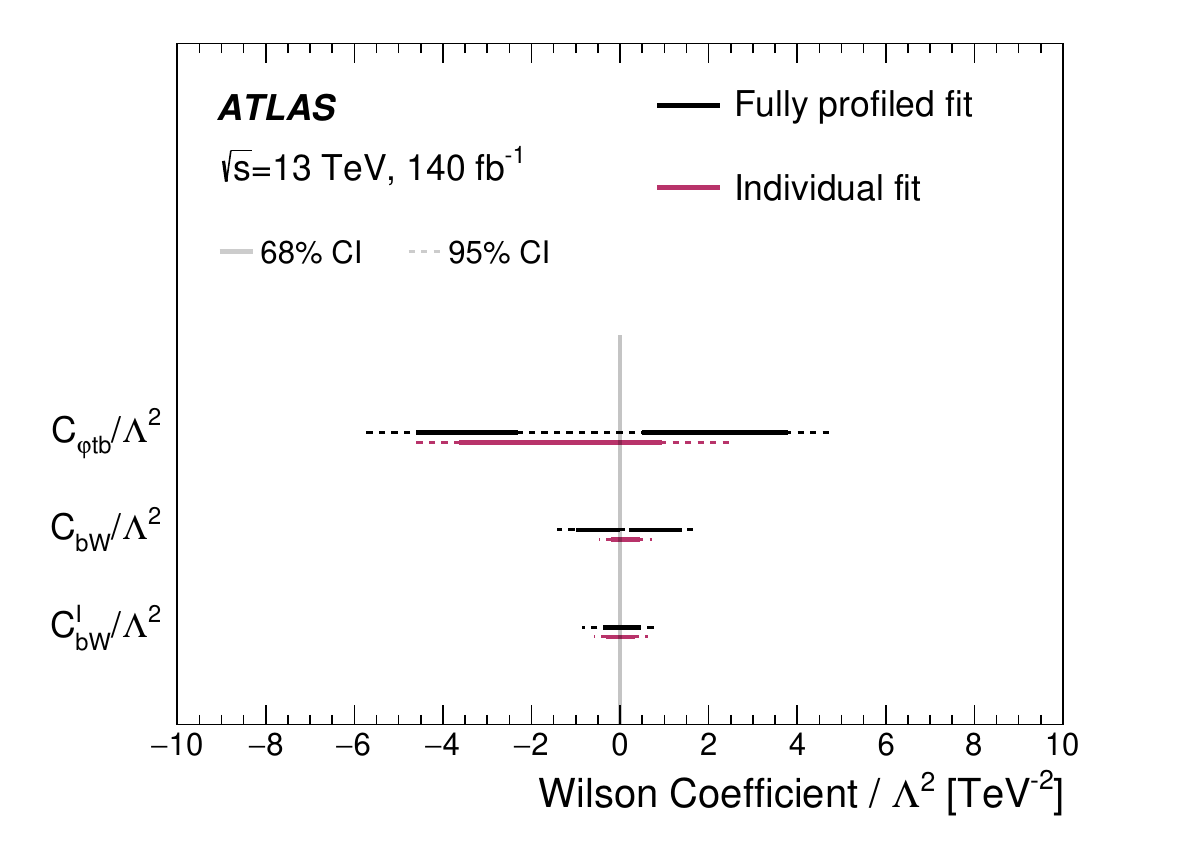}
\caption{}
\end{subfigure}
\caption{Summary of confidence intervals extracted from the likelihood fits described in~\cref{sec:result} for (a) \ctw, \citw, \cqq, \cphiq and (b) \cptb, \cbw, \cibw.
The WC 68\% CL (thick line) and 95\% CL (dotted line)
intervals are shown for each of the different fitting schemes.}%
\label{fig:results_summary}
\end{center}
\end{figure}

The constraints on all seven WCs are summarised in \cref{fig:results_summary}. Competitive constraints are placed on all of these parameters, the most restrictive being those on the real and imaginary parts of $c_{tW}$.
The sensitivity to the various WCs arises from different aspects of the data:
\begin{itemize}
\item The imaginary components of $c_{tW}$ and the product of $c_{\varphi tb}$ and $c_{bW}$ are constrained by the imaginary parts of the \mfunc coefficients, while the constraints on the real parts of $c_{tW}$, $c_{bW}$, and $c_{\varphi tb}$ are determined by the real parts of the \mfunc coefficients. Since $\mathcal{O}_{bW}$ and $\mathcal{O}_{\varphi tb}$ do not produce large interference with the SM amplitude, the corresponding WCs are less constrained.
\item \cphiq and \cqq are constrained by the event counts in the two \pt regions; \cphiq affects the overall rate, whereas \cqq impacts the high-\pt region more than the low-\pt region. Having multiple \pt regions allows one to separate the effects of these two WCs.
A non-zero \cqq also predicts a slight reduction in the polarisation of the produced top quarks, with more of an effect in the high-\pt region~\cite{Brivio:2019ius}.
The fitted result of \cqq, which corresponds to a $2.8\sigma$ deviation from the SM, is attributed to the small difference between the number of events observed in the high-\pt signal region and the number predicted by the \POWHEGBOX[v2] simulation. More studies are
needed to determine whether NNLO effects could provide better agreement with the data. Since both \cqq and \cphiq affect the event counts, two bins in \pt are the minimum required to separate their effects. In the future, it would be beneficial to use more \pt bins
to better determine \cqq.
\end{itemize}

\FloatBarrier

The best previously published constraint on \ctw comes from a measurement of the fraction of top-quark decays that produce each \PW-boson helicity state ($F_0$, $F_{\text{L}}$, $F_{\text{R}}$) in \textit{pp} collisions at $\sqrt{s}=$\SI{8}{\TeV}, assuming that $\cpq=0$~\cite{TOPQ-2018-02}.
This analysis improves on that result by a factor of two, while removing the assumption that the other WCs are zero.
The two-dimensional limit contours for \cptb and \cbw improve on the results from the measurement of $F_{\text{R}}$ in Ref.~\cite{TOPQ-2018-02} by providing constraints along the direction to which $F_{\text{R}}$ is blind, while achieving comparable precision along the direction to which $F_{\text{R}}$ is sensitive.
This result also improves on the limits on \citw from measurements of top-quark polarisation~\cite{TOPQ-2018-10} by a factor of three.
More recent measurements from CMS~\cite{CMS-TOP-22-006} also yield constraints on the WCs investigated here.
The precision of those constraints is a factor of two better in \cptb
and \cbw, and complementary to those presented here; however, they do
not provide any information on the imaginary components of these WCs.
Reference~\cite{CMS-TOP-22-006} also provides tighter constraints on \cqq by a factor of three and shows no significant tension with the SM.
The constraints on \cqq presented here are found to be compatible, within uncertainties, with Ref.~\cite{CMS-TOP-22-006}.
After the submission of this manuscript, a related study was reported
in Ref.~\cite{TOPQ-2023-44}. Although that analysis was not optimised
for the extraction of EFT parameters, it indicates that the inclusion
of NNLO effects improves the agreement between SM predictions and the
data. The measurement of the total \tch single-top-quark
cross-section~\cite{TOPQ-2019-02} provides slightly better constraints
on \cpq, but only under the assumption that all other WCs are zero.
This paper leads to an improvement in the constraints on \cpq by a factor of two.

It should be pointed out that while NLO and EW corrections~\cite{Gonz_lez_Sprinberg_2015,Arhrib_2016,Fischer_2018} have been
calculated for the structure functions, defined in Ref.~\cite{Fischer:2001gp}, to the $M$-function
coefficients used in this analysis, NNLO corrections are currently
available only for those associated with $W$-boson helicity
fractions~\cite{Czarnecki:2010gb}. With additional data from LHC Run 3, together with an
improved understanding of systematic uncertainties, the overall
uncertainty could be reduced to a level comparable to the expected
size of these NNLO corrections.

% End of text imported from the .//sections/results.tex input file

%
%
%
%
\FloatBarrier

\section{Conclusion}
\label{sec:conclusion}

% The next lines are included from the .//sections/conclusions.tex input file

In summary, constraints are placed on dimension-six WCs at the \Wtb\ vertex, and on the four-quark coefficient \cqq, using the \lumi dataset collected from $\sqrt{s}=\SI{13}\xspace\mathrm{TeV}\xspace$ \pp collisions by the ATLAS detector during \RunTwo of the LHC.
The strategy used in this study offers an approach for probing EFT WCs with high precision through Fourier-based angular analysis.
The quadruple-differential decay spectrum of \tch single-top-quark events is projected onto an orthonormal \mfunc basis, and the 23 leading normalised \mfunc coefficients for each of four different channels are fitted simultaneously with the event yields using a maximum-likelihood method.
Although the systematic uncertainty is the dominant source of uncertainty overall, no individual source of systematic uncertainty significantly outweighs the total statistical uncertainty, which is primarily driven by the size of the data sample.
All best-fit points are consistent with the SM expectation.
%

%
%
%
%
%
%
%
%
%
%

% End of text imported from the .//sections/conclusions.tex input file

%
%
%
%
%
%
\section*{Acknowledgements}
%

%

% The next lines are included from the .//acknowledgements/Acknowledgements.tex input file
%
%

%
%

We thank CERN for the very successful operation of the LHC and its injectors, as well as the support staff at
CERN and at our institutions worldwide without whom ATLAS could not be operated efficiently.

The crucial computing support from all WLCG partners is acknowledged gratefully, in particular from CERN, the ATLAS Tier-1 facilities at TRIUMF/SFU (Canada), NDGF (Denmark, Norway, Sweden), CC-IN2P3 (France), KIT/GridKA (Germany), INFN-CNAF (Italy), NL-T1 (Netherlands), PIC (Spain), RAL (UK) and BNL (USA), the Tier-2 facilities worldwide and large non-WLCG resource providers. Major contributors of computing resources are listed in Ref.~\cite{ATL-SOFT-PUB-2026-001}.

We gratefully acknowledge the support of ANPCyT, Argentina; YerPhI, Armenia; ARC, Australia; BMWFW and FWF, Austria; ANAS, Azerbaijan; CNPq and FAPESP, Brazil; NSERC, NRC and CFI, Canada; CERN; ANID, Chile; CAS, MOST and NSFC, China; Minciencias, Colombia; MEYS CR, Czech Republic; DNRF and DNSRC, Denmark; IN2P3-CNRS and CEA-DRF/IRFU, France; SRNSFG, Georgia; BMFTR, HGF and MPG, Germany; GSRI, Greece; RGC and Hong Kong SAR, China; ICHEP and Academy of Sciences and Humanities, Israel; INFN, Italy; MEXT and JSPS, Japan; CNRST, Morocco; NWO, Netherlands; RCN, Norway; MNiSW, Poland; FCT, Portugal; MNE/IFA, Romania; MSTDI, Serbia; MSSR, Slovakia; ARIS and MVZI, Slovenia; DSI/NRF, South Africa; MICIU/AEI, Spain; SRC and Wallenberg Foundation, Sweden; SERI, SNSF and Cantons of Bern and Geneva, Switzerland; NSTC, Taipei; TENMAK, T\"urkiye; STFC/UKRI, United Kingdom; DOE and NSF, United States of America.

Individual groups and members have received support from BCKDF, CANARIE, CRC and DRAC, Canada; CERN-CZ, FORTE and PRIMUS, Czech Republic; COST, ERC, ERDF, Horizon 2020, ICSC-NextGenerationEU and Marie Sk{\l}odowska-Curie Actions, European Union; Investissements d'Avenir Labex, Investissements d'Avenir Idex and ANR, France; DFG and AvH Foundation, Germany; Herakleitos, Thales and Aristeia programmes co-financed by EU-ESF and the Greek NSRF, Greece; BSF-NSF and MINERVA, Israel; NCN and NAWA, Poland; La Caixa Banking Foundation, CERCA Programme Generalitat de Catalunya and PROMETEO and GenT Programmes Generalitat Valenciana, Spain; G\"{o}ran Gustafssons Stiftelse, Sweden; The Royal Society and Leverhulme Trust, United Kingdom; United States of America.

In addition, individual members wish to acknowledge support from CERN: European Organization for Nuclear Research (CERN DOCT); Chile: Agencia Nacional de Investigaci\'on y Desarrollo (ANID FONDECYT reg. 1230987, FONDECYT 1230812, FONDECYT 1240864, Fondecyt 3240661, Fondecyt Regular 1240721); China: Chinese Ministry of Science and Technology (MOST-2023YFA1605700, MOST-2023YFA1609300), National Natural Science Foundation of China (NSFC - 12175119, NSFC 12275265); Czech Republic: Czech Science Foundation (GACR - 24-11373S), Ministry of Education Youth and Sports (ERC-CZ-LL2327, FORTE CZ.02.01.01/00/22\_008/0004632), PRIMUS Research Programme (PRIMUS/21/SCI/017); EU: H2020 European Research Council (ERC - 101002463); European Union: European Research Council (BARD No. 101116429, ERC - 948254, ERC 101089007), European Regional Development Fund (HE COFUND GA No.101081355, ERDF), European Union, Future Artificial Intelligence Research (FAIR-NextGenerationEU PE00000013), Italian Center for High Performance Computing, Big Data and Quantum Computing (ICSC, NextGenerationEU), Marie Sklodowska-Curie Actions (GAP-101168829); France: Agence Nationale de la Recherche (ANR-21-CE31-0013, ANR-21-CE31-0022, ANR-22-EDIR-0002, ANR-24-CE31-0504-01); Germany: Deutsche Forschungsgemeinschaft (DFG - 469666862, DFG - CR 312/5-2); China: Research Grants Council (GRF); Italy: Istituto Nazionale di Fisica Nucleare (ICSC, NextGenerationEU), Ministero dell'Università e della Ricerca (NextGenEU 153D23001490006 M4C2.1.1, NextGenEU I53D23000820006 M4C2.1.1, NextGenEU I53D23001490006 M4C2.1.1, SOE2024\_0000023); Japan: Japan Society for the Promotion of Science (JSPS KAKENHI JP22H01227, JSPS KAKENHI JP22H04944, JSPS KAKENHI JP22KK0227, JSPS KAKENHI JP24K23939, JSPS KAKENHI JP24KK0251, JSPS KAKENHI JP25H00650, JSPS KAKENHI JP25H01291, JSPS KAKENHI JP25K01023); Norway: Research Council of Norway (RCN-314472); Poland: Ministry of Science and Higher Education (IDUB AGH, POB8, D4 no 9722), Polish National Science Centre (NCN 2021/42/E/ST2/00350, NCN OPUS 2023/51/B/ST2/02507, NCN OPUS nr 2022/47/B/ST2/03059, NCN UMO-2019/34/E/ST2/00393, UMO-2022/47/O/ST2/00148, UMO-2023/49/B/ST2/04085, UMO-2023/51/B/ST2/00920, UMO-2024/53/N/ST2/00869); Portugal: Foundation for Science and Technology (FCT); Spain: Agencia de Gestión de Ayudas Universitarias y de Investigación (AGAUR - 2023 BP 00141), Generalitat Valenciana (ASFAE/2022/008), Ministry of Science and Innovation (MCIN \& NextGenEU PCI2022-135018-2, MICIN \& FEDER PID2021-125273NB, RYC2019-028510-I, RYC2020-030254-I, RYC2021-031273-I, RYC2022-038164-I), Ministerio de Ciencia, Innovación y Universidades/Agencia Estatal de Investigaci\'on (PID2022-142604OB-C22); Sweden: Carl Trygger Foundation (Carl Trygger Foundation CTS 22:2312), Swedish Research Council (Swedish Research Council 2023-04654, VR 2021-03651, VR 2022-03845, VR 2022-04683, VR 2023-03403, VR 2024-05451), Knut and Alice Wallenberg Foundation (KAW 2018.0458, KAW 2022.0358, KAW 2023.0366); Switzerland: Swiss National Science Foundation (SNSF - PCEFP2\_194658); United Kingdom: Royal Society (NIF-R1-231091); United States of America: U.S. Department of Energy (ECA DE-AC02-76SF00515), Neubauer Family Foundation.

%
%

% End of text imported from the .//acknowledgements/Acknowledgements.tex input file

%
\clearpage
\printbibliography
\clearpage
% ATLAS Collaboration author list
% Reference date of TOPQ-2020-17 is 2025-08-18
% Author list last updated on date 26-FEB-26
% Data extracted on 26-Feb-2026 for paper reference TOPQ-2020-17
% at 2:10am
 
\begin{flushleft}
\hypersetup{urlcolor=black}
{\Large The ATLAS Collaboration}

\bigskip

\AtlasOrcid[0000-0002-6665-4934]{G.~Aad}$^\textrm{\scriptsize 103}$,
\AtlasOrcid[0000-0001-7616-1554]{E.~Aakvaag}$^\textrm{\scriptsize 17}$,
\AtlasOrcid[0000-0002-5888-2734]{B.~Abbott}$^\textrm{\scriptsize 122}$,
\AtlasOrcid[0000-0002-0287-5869]{S.~Abdelhameed}$^\textrm{\scriptsize 118a}$,
\AtlasOrcid[0000-0002-1002-1652]{K.~Abeling}$^\textrm{\scriptsize 55}$,
\AtlasOrcid[0000-0001-5763-2760]{N.J.~Abicht}$^\textrm{\scriptsize 49}$,
\AtlasOrcid[0000-0002-8496-9294]{S.H.~Abidi}$^\textrm{\scriptsize 30}$,
\AtlasOrcid[0009-0003-6578-220X]{M.~Aboelela}$^\textrm{\scriptsize 45}$,
\AtlasOrcid[0000-0002-9987-2292]{A.~Aboulhorma}$^\textrm{\scriptsize 36e}$,
\AtlasOrcid[0000-0001-5329-6640]{H.~Abramowicz}$^\textrm{\scriptsize 156}$
% ...
\bigskip
\\

% ...
$^{17}$Department for Physics and Technology, University of Bergen, Bergen; Norway.\\
% ...
$^{30}$Physics Department, Brookhaven National Laboratory, Upton NY; United States of America.\\
% ...
$^{36}$$^{(a)}$Facult\'e des Sciences Ain Chock, Universit\'e Hassan II de Casablanca;$^{(b)}$Facult\'{e} des Sciences, Universit\'{e} Ibn-Tofail, K\'{e}nitra;$^{(c)}$Facult\'e des Sciences Semlalia, Universit\'e Cadi Ayyad, LPHEA-Marrakech;$^{(d)}$LPMR, Facult\'e des Sciences, Universit\'e Mohamed Premier, Oujda;$^{(e)}$Facult\'e des sciences, Universit\'e Mohammed V, Rabat;$^{(f)}$Institute of Applied Physics, Mohammed VI Polytechnic University, Ben Guerir; Morocco.\\
% ...
$^{45}$Physics Department, Southern Methodist University, Dallas TX; United States of America.\\
% ...
$^{49}$Fakult\"{a}t Physik, Technische Universit{\"a}t Dortmund, Dortmund; Germany.\\
% ...
$^{55}$II. Physikalisches Institut, Georg-August-Universit\"{a}t G\"ottingen, G\"ottingen; Germany.\\
% ...
$^{103}$CPPM, Aix-Marseille Universit\'e, CNRS/IN2P3, Marseille; France.\\
% ...
$^{118}$$^{(a)}$New York University Abu Dhabi, Abu Dhabi;$^{(b)}$United Arab Emirates University, Al Ain; United Arab Emirates.\\
% ...
$^{122}$Homer L. Dodge Department of Physics and Astronomy, University of Oklahoma, Norman OK; United States of America.\\
% ...
$^{156}$Raymond and Beverly Sackler School of Physics and Astronomy, Tel Aviv University, Tel Aviv; Israel.\\
% ...
$^{a}$ Also at Affiliated with an institute formerly covered by a cooperation agreement with CERN.\\
$^{b}$ Also at An-Najah National University, Nablus; Palestine.\\
$^{c}$ Also at Borough of Manhattan Community College, City University of New York, New York NY; United States of America.\\
$^{d}$ Also at Center for Interdisciplinary Research and Innovation (CIRI-AUTH), Thessaloniki; Greece.\\
$^{e}$ Also at Centre of Physics of the Universities of Minho and Porto (CF-UM-UP); Portugal.\\
$^{f}$ Also at CERN, Geneva; Switzerland.\\
% ...

\end{flushleft}

% Created with Glance <Atlas.Glance@cern.ch>

%

%
%

%

%
%
%
%

\end{document}